\documentclass[fleqn,usenatbib]{mnras}

\usepackage[T1]{fontenc}

\DeclareRobustCommand{\VAN}[3]{#2}
\let\VANthebibliography\thebibliography
\def\thebibliography{\DeclareRobustCommand{\VAN}[3]{##3}\VANthebibliography}

\usepackage{graphicx}	
\usepackage{amsmath}	
\usepackage{amssymb}	
\usepackage{newtxtext,newtxmath}
\usepackage{pdflscape}
\usepackage{physics}
\renewcommand{\arcsec}{$"$}

\renewcommand{\arcsec}{arcsec}

\title[BPMG M-dwarfs with ALMA]{An ALMA Survey of M-dwarfs in the Beta Pictoris Moving Group with Two New Debris Disc Detections}

\author[P. F. Cronin-Coltsmann et al.]{
\parbox{\textwidth}{Patrick F. Cronin-Coltsmann,$^{1,2}$\thanks{E-mail: patrick.cronin-coltsmann@warwick.ac.uk }
Grant M. Kennedy,$^{1,2}$
Quentin Kral,$^{3}$
Jean-Fran\c cois Lestrade,$^{4}$
Sebastian Marino,$^{5}$
Luca Matr\`{a},$^{6}$
Mark C. Wyatt$^{7}$}
\\
\\
\parbox{\textwidth}{
$^{1}$Department of Physics, University of Warwick, Gibbet Hill Road, Coventry, CV4 7AL, UK\\
$^{2}$Centre for Exoplanets and Habitability, University of Warwick, Gibbet Hill Road, Coventry CV4 7AL, UK\\
$^{3}$LESIA, Observatoire de Paris, Universit\'e PSL, CNRS, Sorbonne Universit\'e, Univ. Paris Diderot, Sorbonne Paris Cit\'e, 5 place Jules Janssen, 92195 Meudon, France\\
$^{4}$LERMA, Observatoire de Paris, PSL Research University, CNRS, Sorbonne Universit\'es, UPMC Univ. Paris 06, 75014 Paris, France\\
$^{5}$School of Physics and Astronomy, University of Exeter, Stocker Road, Exeter, EX4 4QL, UK \\
$^{6}$School of Physics, Trinity College Dublin, the University of Dublin, College Green, Dublin 2, Ireland \\
$^{7}$Institute of Astronomy, University of Cambridge, Madingley Road, Cambridge CB3 OHA, UK\\
}}

\date{Accepted 2/10/2023. Received YYY; in original form ZZZ}

\pubyear{2023}

\begin{document}
\label{firstpage}
\pagerange{\pageref{firstpage}--\pageref{lastpage}}
\maketitle

\begin{abstract}

Previous surveys in the far-infrared have found very few, if any, M-dwarf debris discs among their samples. It has been questioned whether M-dwarf discs are simply less common than earlier types, or whether the low detection rate derives from the wavelengths and sensitivities available to those studies. The highly sensitive, long wavelength Atacama Large Millimetre/submillimetre Array can shed light on the problem.
This paper presents a survey of M-dwarf stars in the young and nearby Beta Pictoris Moving Group with ALMA at Band 7 (880\,$\mu$m). From the observational sample we detect two new sub-mm excesses that likely constitute unresolved debris discs around GJ\,2006\,A and AT\,Mic\,A and model distributions of the disc fractional luminosities and temperatures. From the science sample of 36 M-dwarfs including AU\,Mic we find a disc detection rate of 4/36 or 11.1$^{+7.4}_{-3.3}$\% that rises to 23.1$^{+8.3}_{-5.5}$\% when adjusted for completeness. We conclude that this detection rate is consistent with the detection rate of discs around G and K type stars and that the disc properties are also likely consistent with earlier type stars. We additionally conclude that M-dwarf stars are not less likely to host debris discs, but instead their detection requires longer wavelength and higher sensitivity observations than have previously been employed.

\end{abstract}




\section{Introduction}

M-dwarfs are the most abundant type of star in the sky \citep{Ledrew01}, and these stars have a multitude of detected planets \citep[e.g.][]{Bonfils13,Dressing15,Mulders15}. However, when it comes to debris discs M-dwarfs are distinctly lacking. The far-IR Herschel DEBRIS survey detected infrared excesses around 17\% of FGK type stars \citep{Sibthorpe18} and 24\% of A-type stars \citep{Thureau14}, but only detected two excesses around M-types \citep[GJ\,581; Fomalhaut\,C;][]{Lestrade12,Kennedy13} from a sample of 89 stars for a detection rate of 2\%.
There are only eight nearby M-dwarf discs published in the literature. Of these 3 have yet to be fully resolved: GJ\,581 \citep{Lestrade12}, GJ\,433 and GJ\,649 \citep{Kennedy18a}. 
The remaining 5 have been fully resolved: AU\,Mic \citep[][]{MacGregor13,Daley19}, Fomalhaut\,C \citep{Coltsmann21} and GSC\,07396-00759 \citep{Coltsmann21} with ALMA, and AU\,Mic \citep{Kalas04}, TWA\,7 \citep{Choquet16}, TWA\,25 \citep{Choquet16} and GSC\,07396-00759 \citep{Sissa18,Adam21} in scattered light, confirming that the infrared excesses indeed originate from circumstellar discs. These discs are distinguished from so-called Peter Pan discs around some young M-types (e.g. \citet{Silverberg20}) as they do not show the evidence of ongoing accretion that Peter Pan discs do. In the case of Peter Pan discs, this accretion is indicative of a long-lived gas component that may be a primordial remnant of the original protoplanetary disc. 

The low rate of disc detections could be because the discs simply are not there. It is possible that the high incidence of planets around M dwarfs marks a high efficiency of planet formation, limiting leftover material that would constitute a debris disc.
Alternatively photoevaporation \citep{Adams04} and stellar encounters \citep{Lestrade11} could strip material from M star discs that are forming in cluster environments. 
If discs are present, their underlying physical processes are different to discs around earlier type stars. The low host luminosity is not significant enough for radiation pressure to overcome gravity and instead stellar wind becomes a significant force. It is possible that strong stellar wind drag could remove grains quickly enough that the discs dynamics are different, affecting observability \citep{Plavchan09}.

Alternatively, a population of discs similar to that around early type stars could exist around M-dwarfs but remain difficult to detect with far-IR methods. A lower host luminosity would illuminate the same disc less well and heat {\rm it} to a lower temperature, requiring more sensitive, longer wavelength observations than those employed by previous surveys. 
The Atacama Large Millimetre Array is the best suited contemporary telescope to fulfill these requirements. 

\citet{Luppe20} investigate the capability of ALMA to detect a population of M-dwarf discs around the DEBRIS sample of M-stars, assuming that those discs have the same properties as the DEBRIS FGK-type systems. They conclude that for 15 minutes of observation at Band 7 there would be a 4-16\% detection rate if all the discs were unresolved and a detection rate of 1-6\% if some discs are large or close enough to be resolved. If the discs are resolved, the signal per beam would be reduced and/or some flux would be unrecoverable if the angular scale of the disc is larger than the maximum recoverable scale of the observation's interferometry. 

Debris disc detection rate and fractional luminosity is known to decrease with age as material is lost from the system due to the blow out of dust and the collisional depletion of the reservoir of parent planetesimals \citep{Decin03,Rieke05,Trilling08,Kral13,Montesinos16}. For this reason, if a survey were to be optimised to recover as many disc detections as possible, a sample of young stars should be selected. The $\beta$ Pictoris Moving Group (BPMG) is both young \citep[$\sim$20\,Myr,][]{Bell15,Miret20} and nearby \citep[$\lesssim$100\,pc,][]{Shkolnik17}, making it a valuable stellar sample. \citet{Pawellek21} analyse the F-type population of the BPMG with far-IR photometry and ALMA and find a 75\% detection rate, a significantly higher rate than for the old field stars of the DEBRIS F star sample \citep{Sibthorpe18}, further solidifying the BPMG as a good candidate sample to search for new discs. Indeed, already two of the published M-dwarf discs, AU\,Mic and GSC\,07396-00759, are members of the BPMG. 

In this paper we present observations of the BPMG M-dwarf sample with ALMA. The observational details are presented in \S\ref{sec:Obs}. The results of the survey for individual stars of interest is presented in \S\ref{sec:stars} and new disc detections and the context of the detection rate is discussed in \S\ref{sec:Disc}.

\section{Observations}\label{sec:Obs}

\subsection{Observation Sample}

The observation sample of 39 stars was selected in 2017 for ALMA Cycle 5 based on these criteria: the star is identified as a known member from the literature of the BPMG, the star is identified as an M-type, and the star is within ALMA's observable declination range - i.e. between $\sim$ -65$^\circ$ and 40$^\circ$. These sources were used for the sample selection: \citet{Binks16,Malo13,Shkolnik12,Schlieder10,Lepine09,Zuckerman01}. The sample selection was not informed by the previous detection of any infrared excesses and thus the sample is unbiased in this regard. The sample that satisfies these criteria is now significantly larger, e.g. approximately doubling later in 2017 with new members confirmed by \citet{2017AJ....154...69S}. While observing more targets always provides better statistics, our sample is sufficient for our purposes here.

AU\,Mic is a member of the scientific sample used in the analysis but was not chosen to be observed in the survey as it has already been significantly observed with ALMA. Had it been observed, it would definitely have been re-detected and the new re-observation would not significantly build upon previous observations.

The sample was observed under project 2017.1.01583.S, with further details to follow in \S\ref{sec:ObsDet}.
There were 33 individual ALMA observations, of which two contained both stars of a well studied binary within the field of view (HD\,139084\,AB and AT\,Mic\,AB). A further three contained two Gaia DR3 sources with similar parallax measurements {bf that reside} within the field of view (2MASS\,J05241914-1601153, LP\,476-207, GSC\,08350-01924), i.e. these stars are newly resolved by Gaia to have binary companions. These bring the total confirmed BPMG member stars observed by our survey to 38. Two more observations contained a second Gaia DR3 source without a parallax but with an appropriate G magnitude and sub-arcsecond separation from the primary (2MASS\,J19102820-2319486, UCAC3\,124-580676), i.e. these are potential but unconfirmed binary companions; these are not included as separate stars in our analysis and so do not add to our total.
TYC\,7443-1102-1 is listed alternatively as K9IVe \citep{Pecaut13} and M0.0V \citep{Lepine09}, and so was included in this sample and treated as an M-dwarf, it was later noted to have an infrared excess in Herschel PACS \citep{Tanner20}. One of the observed stars, HD\,139084\,A is a K0V, and so is not part of the scientific sample; this means that only 37 of the 38 stars observed in this survey are included in the scientific sample. Adding AU\,Mic brings the scientific sample to a final total of 38 confirmed M-dwarfs to be analysed.

UCAC4\,345-006842 (AKA Karmn J05084-210) was intended to be observed but the ALMA observation was mispointed, so it was not observed. GJ\,3305 (AKA StKM 1-497), GJ\,182 (AKA V1005\,Ori) and TWA\,22 (AKA ASAS\,J101727-5354.4) were intended to be observed with ALMA, but the scheduling blocks were timed out at the end of the observing period. These stars are for these reasons not part of our scientific sample.

Table \ref{tab:Sample} displays details of our sample of stars. Spectral types for this table were taken from SIMBAD \citep{SIMBAD} unless otherwise noted with an asterisk, luminosities are taken from stellar SED models using available photometry and parallaxes unless otherwise noted with an asterisk. For asterisk noted properties we make estimates using the online 'Modern Mean Dwarf Stellar Color and Effective Temperature Sequence' table\footnote{\url{http://www.pas.rochester.edu/~emamajek/EEM_dwarf_UBVIJHK_colors_Teff.txt}} of \citet{Pecaut13}. The spectral type of TYC\,7443-1102-1 marked with two asterisks is derived from \citet{Lepine09}.

\subsection{Observation Details}\label{sec:ObsDet}

All new observations were performed by ALMA Band 7 (0.87\,mm, 345\,GHz) under project 2017.1.01583.S. We anticipated of order ten detections (i.e. many non-detections), so did not aim to also obtain spectral information by observing with more than one band. The observations were spread across configurations C43-1, C43-2, and C43-3 depending on stellar distance to retain sensitivity to a similar physical scale and avoid resolving out disc emission. Observation details for individual sources can be found in Table\,\ref{tab:ObsDet}.

The spectral setup for all observations comprised four windows centred on 347.937, 335.937, 334.042 and 346.042 GHz with bandwidth 2 GHz and 128 channels for all but the last with width 1.875 GHz and 3840 channels. 
The last window was used to search for CO gas via the J=3-2 emission line, which has also been detected in another young debris disc around the M-dwarf TWA\,7 \citep{Matra19}.

The raw data were calibrated with the provided ALMA pipeline script in \textsc{casa} version 5.1.2-4 \citep{CASA}. To reduce the data volume the visibilities were averaged in 30 second intervals and down to two channels per spectral window for the continuum imaging. All images were generated with the \textsc{clean} algorithm in \textsc{casa}.

\begin{landscape}
\begin{table}
\centering

\caption[Stars in our BPMG sample.]{Stars observed in our sample. Spectral types are derived from SIMBAD unless marked with asterisks, luminosities are taken from stellar SED models using available photometry and parallaxes unless otherwise noted with an asterisk. For asterisk noted properties we make estimates using the online temperature sequence table of \citet{Pecaut13}. The spectral type of TYC\,7443-1102-1 marked with two asterisks is derived from \citet{Lepine09}}
\label{tab:Sample}
\begin{tabular}{llllrl}
\hline
Name & Alternative name & Type & Luminosity [$L_{\odot}$] & Distance [pc] & Notes\\
\hline                                     

2MASS\,J05195327+0617258 & GSC2.3 N9OB003170 & M6.5V* &  0.0057 & 96.1 & - \\
2MASS\,J05241914-1601153 AB & PM J05243-1601 AB & M4.5 & 0.043 & 31.1 & GDR3 Binary\\
2MASS\,J19102820-2319486 & 1SWASP J191028.18-231948.0 & M4 & 0.11 & 59.0 & Possible GDR3 Binary\\
2MASS\,J20333759-2556521 & SCR J2033-2556 & M4.5 & 0.0305 & 43.5 & - \\
ASAS J164301-1754.4 & UCAC4\,361-079084 & M0.5 & 0.141 & 71.1 & - \\
Barta 161 12 & UCAC4\,414-001790 & M4.3V & 0.05 & 37.3 & Spectroscopic Binary \\
BD+30 397 B & V* AG Tri B & M0 & 0.078 & 40.9 & Companion to BD+30 397 A\\
CD-57 1054 & GSC\,08513-00572 & M0Ve & 0.174 & 26.9 & -\\
EPIC 211046195 & 2MASSW J0335020+234235 & M8.5V & 0.00402 & 51.2  & -\\
GJ\,2006\,A & ** LDS 18A & M3.5Ve & 0.053 & 35.0 & Companion to GJ\,2006B\\
GJ\,2006\,B & ** LDS 18B & M3.5Ve & 0.0429 & 35.0 & Companion to GJ\,2006A \\
GJ 3076 & LP\,467-16 & M5.93 & 0.008 & 17.2 & -\\
GSC\,07396-00759 & ASAS J181422-3246.2 & M1Ve & 0.135 & 71.4 & Companion to V4046 Sgr \\
GSC\,08350-01924 AB & 1RXS J172919.1-501454 AB & M3V & 0.163 & 62.6 & GDR3 Binary \\
HD\,139084 & CD-57 6042 A & K0V & 0.98 & 39.3 & Companion to HD\,139084 B, Spectroscopic Binary \\
HD\,139084 B & CD-57 6042 B & M5Ve & 0.0203 & 39.3 & Companion to HD\,139084 \\
HD\,155555\,C & V824 Ara C & M3Ve & 0.044 & 30.3 & Companion to HD\,155555 AB\\
L 836-122 & GJ 3832 & M3.5V & 0.015 & 28.6 & -\\
LP\,353-51 & HIP 11152 & M1V & 0.0641 & 27.2  & -\\
LP\,476-207 AB & GJ 3322 AB & M3.5V & 0.07 & 33.2 & GDR3 Binary/Spectroscopic Binary \\
MCC 124 & HIP 50156 & M0.7V & 0.132 & 23.4 & \\
AT\,Mic\,A & GJ 799 A & M4.5Ve  & 0.035 & 9.9 & Companion to AT\,Mic\,B, companion to AU Mic\\
AT\,Mic\,B & GJ 799 B & M4.5Ve  & 0.031 & 9.8 & Companion to AT\,Mic\,A, companion to AU Mic\\
RX\,J0217.9+1225 & PM J02179+1225 & M4 & 0.0593 & 63.1 & -\\
Smethells 20 & TYC\,9073-762-1 & M1Ve & 0.134 & 50.6 & -\\
TYC\,2211-1309-1 & RX\,J2200.7+2715 & M0.0V & 0.0841 & 35.6 & -\\
TYC\,6872-1011-1 & 1RXS J185803.4-295318 & M0Ve & 0.275 & 74.2 & Spectroscopic Binary \\
TYC\,7443-1102-1 & PM J19560-3207 & M0.0V** & 0.154 & 51.3 & Companion to UCAC3\,116-474938 \\
UCAC2\,19527490 & 2MASS\,J18580464-2953320 & M3V* & 0.12 & - & Likely Companion to TYC\,6872-1011-1\\
UCAC2\,20312880 & RX\,J0613.2-2742 & M3.5 & 0.089 & 32.7 & Double star\\
UCAC3\,116-474938 & 2MASS\,J19560294-3207186 & M4 & 0.11 & 51.3 & Companion to TYC\,7443-1102-1, Double star\\
UCAC3\,124-580676 & SCR J2010-2801 & M3.0Ve & 0.11 & 48.0 & Possible Gaia DR3 Binary/Spectroscopic Binary\\
UCAC3\,176-23654 & RX\,J0534.0-0221 & M3 & 0.066 & 34.4 & -\\
V* TX PsA & ** LDS 793 B & M5IVe & 0.0203 & 20.8 & Companion to V* WW PsA\\
V* WW PsA & ** LDS 793 A & M4IVe & 0.0462 & 20.8 & Companion to V* TX PsA \\
AU Mic & HD\,197481 & M1Ve & 0.0962 & 9.7 & Not observed in this project, companion to AT\,Mic\,AB\\
\hline

\end{tabular}

\end{table}
\end{landscape}

\begin{landscape}
\begin{table}
\centering

\caption[ALMA observation details for the BPMG sample.]{ALMA Band 7 observation details for stars observed under project 2017.1.01583.S. Note some stars were observed in multiple configurations and are listed once for each individual observation. MRS is maximum recoverable scale and PWV is mean precipitable waver vapour.}
\label{tab:ObsDet}
\begin{tabular}{lccccccc}
\hline
Name & Integration time [minutes] & No. Antennae & Min-Max baseline [m] & MRS [\arcsec] & Date & PWV [mm] & Calibrators \\
\hline                                     

2MASS\,J05195327+0617258 & 16.13 & 43 & 15.1 - 782.1 & 4.4 & 28.08.18 & 0.3 &  J0552+0313, J0423-0120\\
2MASS\,J05241914-1601153 AB & 14.62 & 43 & 15.1 - 313.7 & 6.6 & 07.07.18 & 0.5 &  J0524-0913, J0522-3627\\
2MASS\,J19102820-2319486 & 14.11 & 45 & 15.1 - 500.2 & 5.3 & 19.05.18 & 0.9 & J1924-2914, J1751+0939\\
2MASS\,J20333759-2556521 & 14.16 & 44 & 15.1 - 483.9 & 5.6 & 06.04.18 & 0.7 &  J2056-3208, J1924-2914\\
2MASS\,J20333759-2556521 & 14.16 & 46 & 15.1 - 500.2 & 5.7 & 04.05.18 & 0.3 &  J2056-3208, J1924-2914\\
ASAS J164301-1754.4 & 14.67 & 45 & 15.1 - 500.2 & 5.3 & 19.05.18 & 0.9 & J1733-1304, J1517-2422\\
Barta 161 12 & 14.70 & 46 & 15.0 - 313.7 & 7.0 & 31.05.18 & 0.8 &  J0141-0928, J0006-0623\\
BD+30 397 B & 30.47 & 44 & 15.1 - 500.2 & 5.6 & 24.08.18 & 1.0 &  J0423-0120, J0238+1636\\
CD-57 1054 & 17.20 & 46 & 15.1 - 313.7 & 7.0 & 04.07.18 & 1.0 &  J0550-5732, J0519-4546, J0506-6109\\
CD-57 1054 & 17.20 & 43 & 15.1 - 440.4 & 7.0 & 12.08.18 & 0.9 &  J0550-5732, J0519-4546, J0506-6109\\
EPIC 211046195 & 21.25 & 49 & 15.1 - 783.5 & 4.3 & 31.08.18 & 0.8 &  J0336+3218, J0510+1800\\
GJ\,2006\,A & 14.61 & 45 & 15.0 - 313.7 & 7.0 & 23.05.18 & 0.3 & J0040-3243, J2258-2758\\
GJ\,2006\,B & 14.61 & 45 & 15.0 - 313.7 & 7.0 & 23.05.18 & 0.3 & J0040-3243, J2258-2758\\
GJ 3076 & 18.20 & 46 & 15.1 - 313.7 & 6.9 & 30.06.18 & 0.7 & J0117+1418, J0006-0623\\
GSC\,07396-00759 & 14.67 & 44 & 15.1 - 483.9 & 5.6 & 06.05.18 & 0.7 &  J1924-2914, J1826-2924\\
GSC\,08350-01924 AB & 15.18 & 47 & 15.0 - 313.7 & 7.0 & 19.05.18 & 0.3 &  J1650-5044, J1717-5155, J1924-2914\\
GSC\,08350-01924 AB & 16.19 & 46 & 15.1 - 500.2 & 5.2 & 19.05.18 & 0.9 & J1650-5044, J1717-5155, J1924-2914\\
HD\,139084 AB & 17.19 & 48 & 15.0 - 313.7 & 7.4 & 18.05.18 & 1.0 & J1524-5903, J1427-4206\\
HD\,155555\,C & 21.25 & 44 & 15.1 - 500.2 & 6.0 & 06.05.18 & 0.8 &  J1703-6212, J1427-4206\\
L 836-122 & 14.67 & 46 & 15.0 - 313.7 & 6.9 & 15.05.18 & 1.1 & J1408-0752, J1337-1257\\
LP\,353-51 & 25.20 & 44 & 15.1 - 500.2 & 5.6 & 24.08.18 & 0.9 & J0423-0120, J0238+1636, J0237+2848\\
LP\,476-207 AB & 17.19 & 44 & 15.1 - 500.2 & 5.4 & 23.08.18 & 0.7 & J0510+1800, J0449+1121\\
MCC 124 & 21.75 & 44 & 15.1 - 500.2 & 6.0 & 06.05.18 & 0.7 &  J1025+1253, J1058+0133\\
AT\,Mic\,AB & 14.65 & 47 & 15.0 - 313.7 & 7.0 & 19.05.18  & 0.3 & J1924-2914, J2056-3208\\
RX\,J0217.9+1225 & 17.70 & 45 & 15.1 - 783.5 & 4.3 & 06.09.18 & 0.6 & J0211+1051, J0006-0623, J0224+0659\\
Smethells 20 & 18.70 & 44 & 15.1 - 782.1 & 4.7 & 26.08.18 \& 27.08.18 & 0.8 & J1834-5856, J1924-2914, J1723-6500\\
TYC\,2211-1309-1 & 24.26 & 46 & 15.1 - 783.5 & 4.2 & 05.09.18 & 0.7 & J2253+1608, J2217+2421, J0006-0623\\
TYC\,6872-1011-1 & 14.62 & 45 & 15.1 - 500.2 & 5.3 & 19.04.18 & 0.9 & J1924-2914, J1751+0939\\
TYC\,7443-1102-1 & 12.61 & 48 & 15.1 - 483.9 & 5.6 & 22.08.18 & 0.8 & J1924-2914, J2056-4714\\ 
UCAC2\,19527490 & 14.62 & 45 & 15.1 - 500.2 & 5.3 & 19.04.18 & 0.9 & J1924-2914, J1751+0939\\
UCAC2\,20312880 & 13.33 & 46 & 15.0 - 313.7 & 7.0 & 24.05.18 & 0.6 & J0536-3401, J0522-3627\\
UCAC2\,20312880 & 14.67 & 47 & 15.0 - 330.6 & 6.0 & 05.06.18 & 0.7 & J0536-3401, J0522-3627\\
UCAC3\,116-474938 & 14.62 & 48 & 15.1 - 483.9 & 5.6 & 22.08.18 & 0.8 & J1924-2914, J2056-4714\\
UCAC3\,124-580676 & 14.62 & 48 & 15.1 - 483.9 & 5.6 & 22.08.18 & 0.8 & J1924-2914, J2056-4714\\
UCAC3\,176-23654 & 12.09 & 43 & 15.1 - 782.1 & 4.4 & 28.08.18 & 0.3 & J0552-0313, J0423-0120\\
V* TX PsA & 14.62 & 46 & 15.0 - 455.5 & 6.9 & 11.05.18 & 0.4 & J2258-2758, J0006-0623 \\
V* WW PsA & 14.62 & 46 & 15.0 - 455.5 & 6.9 & 11.05.18 & 0.4 & J2258-2758, J0006-0623  \\

\hline

\end{tabular}

\end{table}
\end{landscape}

\begin{figure*}
        {\includegraphics[width=0.9\textwidth]
        {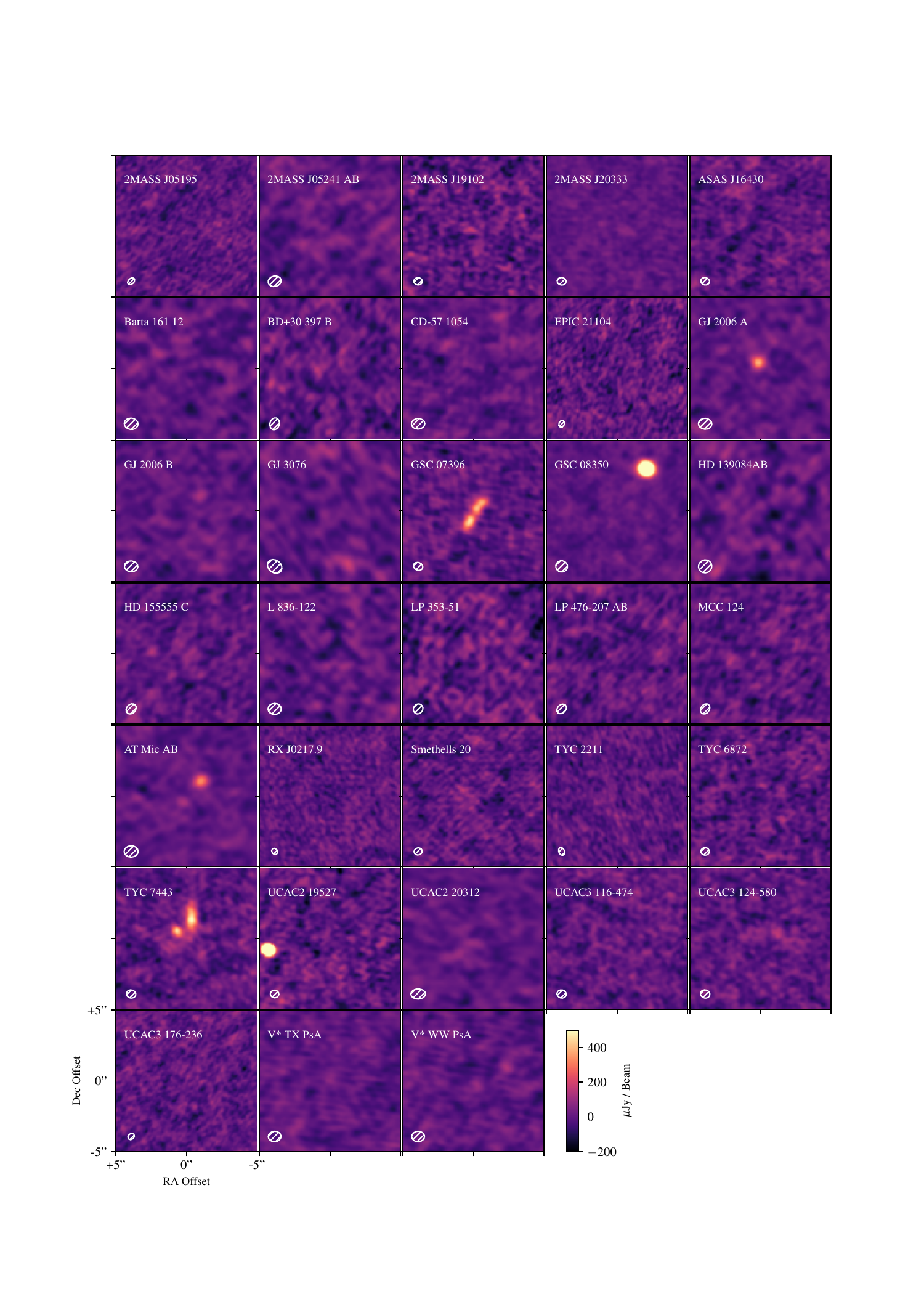}}
        \caption[Naturally weighted ALMA images of the BPMG M-dwarf sample.]{\label{fig:AllObs} Naturally weighted ALMA 880$\mu$m images of our BPMG M-dwarf sample. For all observations except for BD+30 397B and HD\,139084 AB, the star is within 2 arcseconds of the centre of the image. The ellipses in the lower left corners show the restoring beams. }
\end{figure*}

\begin{landscape}
\begin{table}
\centering

\caption[BPMG sample observational results.]{Sample observational results for 880$\mu$m ALMA observations. RMS is the ALMA image root mean square noise as taken from a region surrounding the GAIA DR3 expected stellar location. Beam size is the major-axis of the observation beam. Systems where the on-sky separation is less than the beam size are listed as one with the mean flux of the two \textit{uvmodelfit} fits. Sources in bold have significant excess detections. Parameters for GSC\,07396-00759 are taken from \citet[][]{Coltsmann22} and the radius measurement for AU\,Mic is taken from \citet{MacGregor13}, the expected stellar emission and an 880$\mu$m flux for AU\,Mic are estimated from a combined dust and stellar SED model.}
\label{tab:results}
\begin{tabular}{lcccccccc}
\hline
Name & RMS [$\mu$Jy\,beam$^{-1}$] & Stellar Flux [$\mu$Jy] & Signal [$\mu$Jy\,beam$^{-1}$] & Beam Semi-Major Axis [\arcsec] & Disc radius [au]\\
\hline                                     

2MASS\,J05195327+0617258 & 40 & 0.1 & -32 & 0.573 & - \\
2MASS\,J05241914-1601153 AB & 43 & 8 & 28  & 0.939 & - \\
2MASS\,J19102820-2319486 & 50 & 5 & -9 & 0.607 & - \\
2MASS\,J20333759-2556521 & 23 & 3 & 41 & 0.640 & - \\
ASAS J164301-1754.4     & 47 & 3 & 0.6 & 0.626 & - \\
Barta 161 12            & 40 & 6 & -0.5 & 0.985 & -  \\
BD+30 397 B             & 85 & 6 & 10 & 0.853 & - \\
CD-57 1054              & 40 & 2 & 26 & 0.954 & - \\
EPIC 211046195          & 46 & 0.5 & -87 & 0.515 & - \\
\textbf{GJ\,2006\,A}    & 33 & 6 & 390 & 0.958 & $<$34 \\
GJ\,2006\,B             & 33 & 6 & 2 & 0.957 & - \\
GJ 3076                 & 36 & 6 & 38 & 1.110 & -   \\
\textbf{GSC\,07396-00759}& 40 & 2 & 1840 & 0.683 & 70  \\
GSC\,08350-01924 AB     & 25 & 6 & -12 & 0.840 & - \\
HD\,139084              & 60 & 20 & 133 & 0.960 & - \\
HD\,139084 B            & 60 & 10 & -22 & 0.960 & - \\
HD\,155555\,C           & 40 & 9 & 93 & 0.792 & - \\
L 836-122               & 45 & 3 & -60 & 0.938 & -  \\
LP\,353-51              & 57 & 8 & 19 & 0.734 & -  \\
LP\,476-207 A           & 45 & 20 & 28 & 0.772 & - \\
LP\,476-207 B           & 45 & - & 42 & 0.772 & - \\
MCC 124                 & 45 & 20 & 6 & 0.785 & -  \\
\textbf{AT\,Mic\,A}     & 27 & 70 & 319 & 0.994 & $<$10 \\
AT\,Mic\,B              & 27 & 60 & 120 & 0.994 & - \\
RX\,J0217.9+1225         & 37 & 2 & -12 & 0.485 & - \\
Smethells 20            & 47 & 5 & 75 & 0.582 & -  \\
TYC\,2211-1309-1        & 37 & 5 & -4 & 0.568 & -  \\
TYC\,6872-1011-1        & 47 & 4 & -35 & 0.606 & -  \\
TYC\,7443-1102-1        & 47 & 5 & - & 0.670 & - \\
UCAC2\,19527490         & 50 & 3 & -28 & 0.606 & - \\
UCAC2\,20312880         & 33 & 10 & 39 & 1.042 & - \\
UCAC3\,116-474938       & 40 & 6 & 80 & 0.671 & - \\
UCAC3\,124-580676       & 47 & 7 & 7 & 0.679 & - \\
UCAC3\,176-23654        & 40 & 7 & 25 & 0.519 & - \\
V* TX PsA               & 30 & 8 & 36 & 0.908 & -\\
V* WW PsA               & 35 & 20 & 78 & 0.908 & - \\
\textbf{AU Mic}         & - & 110 & 13000  & - & 40 \\
\hline

\end{tabular}

\end{table}
\end{landscape}

\subsection{Initial image analysis}

Figure \ref{fig:AllObs} shows naturally weighted images of the observational sample generated with the \textsc{clean} algorithm in \textsc{casa}. 
The sample was also visually inspected with 1 and 2\,\arcsec\,$uv$ tapers to search for extended emission. To extract photometry point source models were fit to the visibilities using the \textsc{casa} \textit{uvmodelfit} task at each Gaia DR3 stellar location. We do not allow the offset parameters to vary in these fits to avoid fitting to nearby non-stellar point sources except in the cases of detections and near detections as discussed in \S\ref{sec:stars}.
Fluxes derived from the \textit{uvmodelfit} task are consistent with fluxes measured directly from the images.
The results of these fits and the image parameters can be found in Table\,\ref{tab:results}. Stellar fluxes are estimated by fitting model atmospheres to photometry as outlined in \citet{Yelverton19}; this method uses synthetic photometry of PHOENIX \citep{2013A&A...553A...6H} and blackbody disc models, and \texttt{multinest} \citep{2009MNRAS.398.1601F}, to derive best-fit star and disc parameters. In this table Gaia DR3 confirmed binaries have been split into their individual components with flux measurements taken at the expected location of each component; significant detections are highlighted in bold; parameters for GSC\,07396-00759 are taken from \citet[][]{Coltsmann22}; the radius measurement for AU\,Mic is taken from \citet{MacGregor13}; the expected stellar emission and an 880$\mu$m flux for AU\,Mic are estimated from a combined dust and stellar SED model.

Serendipitous sources within 10\,\arcsec\,of the phase centre whose flux reached at least 5$\sigma$ were identified in the primary beam-corrected \textsc{clean} images and are presented in Table \ref{tab:BGSources}. Sources are identified in ten of the fields. Two sources are present in the TYC\,7443-1102-1 field, one of which is resolved to be 2\,\arcsec\,along one axis. The sources are not associated with any stars and so are likely to be background galaxies. 
The galaxy number count model of \citet{Popping20} can be used to estimate the expected number of galaxies with a flux of at least 0.5 mJy\,beam$^{-1}$ to be present within a 10\,\arcsec\,radius of the phase centre of 33 observations. The expected number of background sources is $12^{+4}_{-10}$, consistent with our detections.

\begin{table*}

\caption[BPMG background sources.]{Background sources. RMS is local to the background source. Fluxes for the TYC\,7443-1102-1 sources noted with an * are integrated fluxes with units $\mu$Jy.}
\label{tab:BGSources}
\begin{tabular}{lcccc}
\hline
Observation & RMS [$\mu$Jy\,beam$^{-1}$] & Source flux [$\mu$Jy\,beam$^{-1}$] & Source Ra [hr:min:sec] & Source Dec [$^\circ$.$^\prime$.$^{\prime\prime}$] \\

\hline                                     
2MASS\,J20333759-2556521 & 40 & 600 & 17:29:20.474 &  -50.14.51.117\\
GSC\,08350-01924 & 25 & 1600 & 0:33:36.964 &  25.57.03.591 \\
Barta 161 12 & 90 & 1600 & 1:35:14.759 & -7.12.52.259  \\
LP\,353-51 & 110 & 800 & 02:23:26.601 & 22.43.54.846 \\
TYC\,2211-1309-1 & 80 & 650 & 22:00:41.823 & 27.15.20.179 \\
TYC\,7443-1102-1 & 47 & 2200* & 19:56:04.396 & -32.07.37.640 \\
TYC\,7443-1102-1 & 47 & 440* & 19:56:04.474 & -32.07.38.475 \\
UCAC2\,19527490 & 65 & 3000 & 18:58:05.016 & -29.53.33.824 \\
UCAC2\,20312880 & 55 & 760 & 06:13:13.748 & -27.41.59.131 \\
UCAC3\,116-474938 & 85 & 800 & 9:56:03.108 & -32.07.29.08 \\
V* TX PsA & 60 & 1300 & 22:44:59.826 & -33.15.32.550 \\
\hline
\end{tabular}

\end{table*}

Significant flux at the stellar location is measured for GJ\,2006\,A, GSC\,07396-00759, AT\,Mic\,A and AT\,Mic\,B, and TYC\,7443-1102-1. GSC\,07396-00759 shows a clearly resolved edge-on disc. The flux from TYC\,7443-1102-1 cannot be differentiated from the background confusion close to the stellar location and so this source is considered significantly confused with no local flux measurement able to be taken. These sources are discussed in more detail in \S\ref{sec:stars}.

Where significant flux is measured at the stellar location we check the observations for signs of mm stellar flares, as these can be mistaken for debris discs \citep[e.g.][]{Anglada17,MacGregor18a}. The observations were split into their individual scans and re-imaged to check for variance of the flux along the time baseline of the observations. No evidence for flaring was found. 

The $^{12}$CO J=3-2 transition line was also checked in these observations by producing \textsc{clean} continuum-subtracted images with the \textit{uvcontsub} algorithm in \textsc{casa} and searching for significant emission at the stellar location and around the expected stellar radial velocity. No CO emission was found in any observation.

A stacked image was also made from the non-detections in which the star is expected to lay within 0.5\,\arcsec\,of the phase centre. With this criterion 2MASS\,J05241914-1601153 AB, BD+30 397 B, GJ\,2006\,B, HD\,139084 B, LP\,476-207 AB, UCAC2\,19527490, UCAC2\,20312880 and UCAC3\,124-580676 are excluded. We also exclude TYC\,7443-1102-1 due to its confusion. The stacked image is thus constituted of the remaining 21 observations and has an RMS of 1$\sigma$ = 10\,$\mu$Jy\,/\,beam. The mean expected stellar emission is 6\,$\mu$Jy\,beam$^{-1}$. No significant flux is found at the centre of the stacked image with a measurement of 12\,$\mu$Jy\,/\,beam, the 3$\sigma$ upper limit on the mean flux for these non-detections is thus 30\,$\mu$Jy\,/\,beam, and the 3$\sigma$ upper limit on mean flux \textit{excess} above the stellar flux is 24\,$\mu$Jy\,/\,beam which at a mean distance of 44 pc corresponds to a disc 25 times less bright than AU\,Mic.

\section{Results}\label{sec:stars}

\subsection{Gaia DR3 parallaxes and binary implications}

The third data release of the Gaia satellite \citep{Gaia22} has improved our astrometric knowledge of our candidate sample since both the proposal submission and observations. Some stars now have accurate parallaxes where there was none before, and other stars have been resolved as binaries with new measurements of their separation.
Multiplicity can cause errors in astrometric solutions \citep{Lindegren18} and this is possibly the root cause for previous difficulty in finding accurate parallaxes.
A measure for non-standard uncertainty in Gaia observations is the astrometric excess noise, \texttt{astrometric\_excess\_noise} (epsi), representing modelling errors and measuring the disagreement between observations of the source and its best fitting model expressed as an angle in units of milli-arcseconds\footnote{\url{https://gea.esac.esa.int/archive/documentation/GDR3/Gaia_archive/chap_datamodel/sec_dm_main_source_catalogue/ssec_dm_gaia_source.html}}.
The epsi in an ideal case should be zero, but for reference the median excess noise for sources with six-parameter solutions is 0.169\footnote{\url{https://gea.esac.esa.int/archive/documentation/GDR3/Data_processing/chap_cu3ast/sec_cu3ast_quality/ssec_cu3ast_quality_properties.html}}. 
A related parameter is the significance of the astrometric excess noise, \texttt{astrometric\_excess\_noise\_sig} (sepsi), for which a value greater than two indicates that the epsi is significant, i.e. the observations of the star significantly differ from its best fitting model. The epsi, when guided by the sepsi, can be used to infer the presence of companions \citep[e.g.][]{Groenewegen18,Kervella19}.

Multiplicity can also affect the likelihood a system contains a detectable debris disc; enhanced collisional evolution from gravitational perturbations can cause the disc flux to decrease more rapidly, so regardless of whether a disc is completely destroyed or not, the disc becomes harder to detect. Empirically, we are always limited by the sensitivity of our observations, so refer to ``detection'' rather than ``existence''. \citet{Yelverton19} find that disc detection rate is more than halved in comparison to single stars when binary separation is less than 25\,au, that the disc detection rate is zero when the separation is between 25 and 135\,au, and that larger separations do not affect disc detection rates. However, the systems studied in that paper were for the majority sun-like, and while a small number of M-type systems were included, the conclusion for sun-like stars might not extend to M-types. All binaries in the sample are now discussed below.

\subsubsection{2MASS\,J05241914-1601153 AB}

2MASS\,J05241914-1601153 (AKA PM J05243-1601, UCAC4\,370-008199) has previously been noted as a double star \citep[][]{Messina17,Miret20} and did not have an accurate parallax prior to Gaia DR3. A has Gaia G magnitude of 12.496$\pm$0.004 and B has a magnitude of 12.778$\pm$0.004, so the stars are of a similar brightness and type. A has a parallax of 32.06$\pm$0.80\,mas and B has a parallax of 32.27$\pm$0.14\,mas placing the stars at 31.1 pc and consistent with co-planarity in the plane of the sky, this would equate their separation of 0.37\,\arcsec\,at the time of observation to 11.5 au. This separation would reduce the likelihood of there being a detectable disc; if a disc is present there is the possibility that it would be circumbinary, which would be resolved by our observations.

\subsubsection{2MASS\,J19102820-2319486}

2MASS\,J19102820-2319486 (AKA 1SWASP J191028.18-231948.0, EPIC 215900519) did not have a parallax measurement prior to Gaia DR3, but now has a measured parallax of 17.0$\pm$0.2\,mas, putting it at 59 pc. \citet{Messina17} label it as a single star, however Gaia DR3 also revealed a second source at a 0.3\,\arcsec\,separation without a parallax or proper motion but with a G magnitude of 12.882$\pm$0.006 compared to 2MASS\,J19102820-2319486's magnitude of 12.528$\pm$0.004. 
The excess astrometric noise for both sources is moderate. The excess astrometric noise is 1.394 mas and the significance of astrometric noise value is 1390 for the source with parallax and the epsi is 2.198 mas and the sepsi is 1900 for the source without parallax. This could explain the lack of a previous Gaia fit for 2MASS\,J19102820-2319486 and the lack of a Gaia fit for the second source.
Multiplicity can be a cause of astrometric noise, and so it is possible the two sources indeed constitute a binary, if approximately in the plane of the sky the separation would be 18 au. This separation would reduce the likelihood of there being a detectable debris disc around either star and any disc could be circumbinary if present.

\subsubsection{Barta 161 12}

Barta 161 12 (AKA UCAC4\,414-001790, ASAS J013514-0712.9, 2MASS\,J01351393-0712517) has parallax $26.82 \pm 0.05$\,mas and distance 37.3\,pc. It is listed as a double-lined spectroscopic binary by \citep{Malo14} and Gaia DR3 detects only one star. Assuming a resolution limit of 0.5\,\arcsec the binary separation is likely less than 19\,au, which would reduce the likelihood of there being a detectable disc and any disc present would likely be circumbinary.

\subsubsection{BD+30 397 B}

BD+30 397 B (AKA 2MASS\,J02272924+3058246, GSC\,02323-00566, AG Tri B) is a companion to the disc hosting star BD+30 397 A \citep[AG Tri, ][]{2008ApJ...681.1484R}. The pair's parallax ($24.42 \pm 0.02$ and $24.43 \pm 0.03$\,mas for A and B respectively, at 40.9\,pc) is consistent with them being co-planar in the plane of the sky and their separation of 22.2\,\arcsec\,equates to 910 au. Their separation is unlikely to affect the likelihood of there being a detectable disc around either star.

BD+30 397 B has a high noise in Table \ref{tab:results} as the observation was pointed near the centre of the binary, placing BD+30 397 B at the edge of the primary beam, raising the local noise. Despite this pointing, BD+30 397 A is outside the 12\,\arcsec\,FWHM of the primary beam, and as such is unobserved.

\subsubsection{GJ\,2006\,AB}

GJ\,2006\,AB (AKA LDS 18A, 2MASS\,J00275023-3233060, UCAC3\,115-1206) have parallax ($28.55 \pm 0.04$ and $28.59 \pm 0.04$\,mas, 35\,pc) consistent with being approximately co-planar in the plane of the sky and their separation of 17.9\,\arcsec\,equates to 625 au. Their separation is unlikely to affect the likelihood of there being a detectable disc around either star.

\subsubsection{GSC\,07396-00759}

GSC\,07396-00759 (AKA ASAS J181422-3246.2, CAB 25B, UCAC4\,287-163100) has parallax $13.92 \pm 0.02$\,mas and distance 71.8\,pc. As noted in \citet[][]{Coltsmann22}, it is a wide separation companion of the well-studied close-binary V4046\,Sgr at a distance of 12,300\,au \citep{Torres06,Kastner11}. V4046\,Sgr possesses both a gas-rich circumbinary disc and evidence of ongoing accretion \citep[e.g.][]{Stempels04,Oberg11,Rosenfeld13,Rapson15,Kastner18,Dorazi19,Brunner22}. The 12,300\,au separation is unlikely to affect the likelihood of there being a detectable disc around either system.

\subsubsection{GSC\,08350-01924 AB}

GSC\,08350-01924 (AKA 1RXS J172919.1-501454, UCAC2\,10274954) has been listed as a binary in previous works \citep{Alonso15,Messina17} and \citet{Zuniga21} conclude it not to be a spectroscopic binary. Gaia DR3 has resolved the binary and identified parallaxes for each star for the first time. A has a parallax of 16.15$\pm$0.06\,mas and B has a parallax of 15.95$\pm$0.078\,mas putting the binary at 62.3\,pc \citep{Bailer21}. 
The difference in parallax of the pair, 0.2$\pm$0.098\,mas, is within two sigma of zero, so if the two are approximately co-planar in the plane of the sky, the binary separation would be 44\,au. A has a Gaia G magnitude of 12.295$\pm$0.003 and B has a magnitude of 12.573$\pm$0.003, so the stars are of a similar brightness and type. If they are widely separated, their separation would be unlikely to affect the likelihood of there being a detectable disc around either star. If they are separated by 44\,au, their separation would make it unlikely that the system hosts a debris disc.

\subsubsection{HD\,139084 AB}

HD\,139084 AB (AKA CD-57 6042 AB, 2MASS\,J15385757-5742273 AB) have parallax measurements of 25.8$\pm$0.2\,mas and 25.55$\pm$0.02\,mas respectively and are separated by 10.3\,\arcsec\,on the sky. The stars therefore constitute a wide binary with a likely separation of at least 50,000\,au. Their separation is unlikely to affect the likelihood of there being a detectable disc around HD\,139084 B, although HD\,139084 A is known to be a single lined spectroscopic binary \citep{Nielsen16} which would reduce its likelihood of hosting a detectable disc. 

HD\,139084 AB have a higher noise in Table \ref{tab:results} as the observation was pointed at the centre of the binary, placing both stars at the edge of the primary beam, raising the local noise.

\subsubsection{HD\,155555\,C}

HD\,155555\,C (AKA V824 Ara C, UCAC3\,47-295205, 2MASS\,J17173128-6657055) is companion to the short period binary HD\,155555 AB with a separation on the sky of 34\,\arcsec; at a distance of 30.3\,pc (parallaxes of $32.95 \pm 0.02$ and $32.88 \pm 0.03$\,mas for AB and C respectively) this equates to a separation on the sky of 1000\,au. Their separation is unlikely to affect the likelihood of there being a detectable disc around either component.

\subsubsection{LP\,476-207 AB}

LP\,476-207 (AKA HIP 23418, GJ 3322, 2MASS\,J05015881+0958587) is a literature double lined spectroscopic binary \citep{Delfosse99} with an orbital period of 11.9 days \citep{Messina17}. Gaia DR3 resolves two stars, we will label LP\,476-207 AB as these two separated components, making the spectroscopic binary LP\,476-207\,AaAb (or possibly BaBb). A has a parallax of 42.04$\pm$0.03\,mas and B has a parallax of 42.10$\pm$0.09\,mas, thus the two are consistent with being approximately co-planar in the plane of the sky. A has a G magnitude of 10.568$\pm$0.003 and B has a magnitude of 11.420$\pm$0.004, thus A is likely the primary and dominates the flux from the system.
Their separation of 1.4\,\arcsec\,on the sky at 33.2 pc equates to 46.5\,au. This separation would make it unlikely that the system hosts a debris disc.

\subsubsection{AT\,Mic\,AB}

AT\,Mic (AKA GJ 799 , HD\,196982, HIP 102141, CD-32 16135, 2MASS\,J20415111-3226073) is a literature close binary system and is highly likely to be a distant companion to AU Mic \citep{Humason27,Caballero09,Shaya11,Messina16} with an on-sky separation of 0.23 pc which equates to 47,000\,au on the sky. The AT\,Mic\,AB binary have Gaia G magnitudes of 9.576$\pm$0.003 and 9.605$\pm$0.003 respectively, so the stars are of a similar brightness and type. The system has been observed to show significant evidence of proper motion \citep[][and references therein]{Messina16} and \citet{Malkov12} provide an orbital period of 209\,yr with a semi-major axis of 3.18\,\arcsec, corresponding to 31\,au, and an eccentricity of $e=0.26$ for the binary.

Gaia DR3 measures the parallaxes for the AT\,Mic binary of 100.79$\pm$0.07\,mas and 101.97$\pm$0.08\,mas, which would be inconsistent with the two being approximately co-planar in the plane of the sky, equating to a separation of 23,300\,au. However, the Gaia DR3 observations for AT\,Mic\,A have an excess astrometric noise of 0.509\,mas and a significance of astrometric noise value of 330, and AT\,Mic\,B has values of 0.502\,mas and 311 respectively. For comparison, their wide separation companion AU Mic has values of 0.098\,mas and 6.1 respectively. The level of astrometric noise is significant and could mean that the uncertainty of the Gaia parallaxes is underestimated. 

Given the extensive historic observation of the system, observed apparent orbital motion and high excess astrometric noise on the Gaia parallaxes, it is likely that the Gaia parallaxes for this system are untrustworthy. Thus, we will continue with the understanding that the stars are co-planar and so are separated primarily by the 2\,\arcsec\,on the sky. Using \citet{Malkov12}'s orbital parameters the semi-major axis of the binary is 31\,au.

The separation with AU\,Mic would be unlikely to affect the likelihood of either system hosting a detectable disc, but the AT\,Mic binary separation would make it unlikely that the system hosts a debris disc.

\subsubsection{TYC\,6872-1011-1 and UCAC2\,19527490}

TYC\,6872-1011-1 (AKA 1RXS J185803.4-295318, UCAC4\,301-253452, 2MASS\,J18580415-2953045) is reported as a double lined spectroscopic binary in \citet{Zuniga21}. The parallax is $13.45 \pm 0.04$\,mas, giving a distance of 74.3\,pc.
The binary separation is likely less than 25\,au as the radial velocity observations were only a few nights apart; this would reduce the likelihood that the system hosts a detectable disc and any disc could be circumbinary.

UCAC2\,19527490 (AKA 2MASS\,J18580464-2953320) does not have a reported parallax in either the literature or Gaia DR3. Gaia DR3 measures a very large excess astrometric noise, the epsi is 59 mas and the sepsi is 240,000, which could be indicative of a close binary companion. A close companion would reduce the likelihood that the system hosts a detectable disc and any disc could be circumbinary. 

UCAC2\,19527490 is only separated from TYC\,6872-1011-1 by 28.3" on the sky, and the two share very similar proper motions and radial velocities, and so it has been posited before that the two are companions \citep{Moor13}. This would place UCAC2\,19527490 at 74.2 pc alongside TYC\,6872-1011-1 and their separation would equate to 2100 au. This separation would not reduce the likelihood of either star hosting a detectable disc.

\subsubsection{TYC\,7443-1102-1 and UCAC3\,116-474938}

TYC\,7443-1102-1 (AKA 2MASS\,J19560438-3207376, PM J19560-3207, UC 4054A) and UCAC3\,116-474938 (AKA 2MASS\,J19560294-3207186, BWL 53) are known to be companions. The two have parallaxes of 19.49$\pm$0.02\,mas and 19.5$\pm$0.7\,mas respectively, consistent with being approximately co-planar in the plane of the sky. At a distance of 51.3 pc their separation of 26.3\,\arcsec\,equates to 1350 au. This separation would not reduce the likelihood of either star hosting a detectable disc.

UCAC3\,116-474938 is also listed as a literature double star \citep{Messina17}. This binarity is not resolved by Gaia DR3 but the star does have a high excess astrometric noise. The epsi is 5.59 mas and the sepsi is 4000, indicating the possible presence of a close companion. A close companion would reduce the likelihood of the system hosting a detectable disc. 

\subsubsection{UCAC2\,20312880}

UCAC2\,20312880 (AKA RX\,J0613.2-2742, 2MASS\,J06131330-2742054) is a literature double star \citep{Messina17} with parallax $29.6 \pm 0.2$\,mas and distance 33.8\,pc. This is not resolved by Gaia DR3 but the star has a high excess astrometric noise, the epsi is 2.5 mas and the sepsi is 960, indicating the possible presence of a close companion. A close companion would reduce the likelihood of the system hosting a detectable disc.

\subsubsection{UCAC3\,124-580676}

UCAC3\,124-580676 (AKA SCR J2010-2801, 2MASS\,J20100002-2801410) is a literature spectroscopic binary and is listed as types M2.5+M3.5 in \citet{Messina17}. Gaia DR3 resolves two stars at a 1\,\arcsec\,separation with primary parallax $21.5 \pm 0.3$\,mas (46.5\,pc) but without a parallax for the secondary. The two stars have Gaia magnitudes of 12.449$\pm$0.005 and 12.207$\pm$0.004 indicating that the two are of similar type. The excess astrometric noise for the sources is very high, the epsi is 2.02 mas and the sepsi is 490 for the source with parallax and the epsi is 14.2 mas and the sepsi is 7360 for the source without parallax, explaining the lack of fit for the secondary. If approximately in the plane of the sky the separation would be 48 au. This separation would make it unlikely that the system hosts a debris disc.

\subsubsection{TX PsA and WW PsA}

TX PsA (AKA GJ 871.1 B, UCAC2\,17853886, 2MASS\,J22450004-3315258 ) and WW PsA (AKA CD-33 16206, GSC\,07501-00987, HIP 112312, 2MASS\,J22445794-3315015) are known companions. Their Gaia DR3 parallaxes are 48.00$\pm$0.03\,mas and 47.92 $\pm$0.03\,mas respectively. \citet{Bailer21} measure distances of 20.826$\pm$0.013\,pc and 20.843$\pm$0.012\,pc respectively, so the stars could be but are not necessarily approximately co-planar in the plane of the sky. The stars are separated in the plane of the sky by 36\,\arcsec; at a distance of 20.8\,pc this equates to 750 au. This separation would not reduce the likelihood of either star hosting a detectable disc.

\subsubsection{Binaries summary}

As it is not an M-star, HD\,139084\,A is excluded from the below summary. Where the parallax measurements of each star in a binary are consistent with each other, we assume that the two stars have equal parallaxes in our analysis. There of course remains the possibility that there is a non-zero separation along the line of sight and so the following separations are strictly speaking minimum possible separations.

One system is a Gaia DR3 resolved binary with both parallaxes and a separation of less than 25\,au (2MASS\,J05241914-1601153\,AB, this separation is less than the observation beam size).
One system is a Gaia DR3 resolved binary with one parallax and a potential separation of less than 25 au (2MASS\,J19102820-2319486, this separation is less than the observation beam size).
Two stars are spectroscopic binaries with no resolved companions in Gaia DR3 (Barta 161 12, TYC\,6872-1011-1).
Two stars are literature double stars unresolved in Gaia DR3 but with high excess astrometric noises (UCAC2\,20312880, UCAC3\,116-474938).
One star is not previously listed as a multiple star but has very high excess astrometric noise (UCAC2\,19527490).
In total there are six (seven if 2MASS\,J05241914-1601153\,AB is counted) systems with a binary separation less than 25\,au; these are half as likely to possess detectable debris discs than single stars, assuming that the results of \citet{Yelverton19} extend to M type stars.

One star is a spectroscopic binary and has two stars resolved in Gaia DR3 with one parallax and a potential separation between 25 and 135 au (UCAC3\,124-580676).
One system is a spectroscopic binary and has two stars resolved in Gaia DR3 with both parallaxes and a separation between 25 and 135 au (LP\,476-207 AB).
One system is a binary and has two stars resolved in Gaia DR3 with both parallaxes (that likely have underestimated uncertainties), has literature orbital parameters and a separation between 25 and 135 au (AT\,Mic\,AB, this separation is greater than the observation beam size).

In total there are three systems with a binary separation between 25 and 135 au that are very unlikely to possess detectable debris discs, assuming that the results of \citet{Yelverton19} extend to M type stars.

Four of the above stars are also companions to other stars with a separation greater than 135 au (UCAC2\,19527490, UCAC3\,116-474938, AT\,Mic\,AB)

A further 9 stars are Gaia DR3 resolved companions to other stars with all parallaxes and a separation greater than 135 au (BD+30\,397\,B, GJ\,2006\,A, GJ\,2006\,B, GSC\,07396-00759, HD\,139084\,B, HD\,155555\,C, TYC\,7443-1102-1, TX\,PsA, WW\,PsA). The multiplicity of these stars is unlikely to affect the likelihood of the presence of a detectable debris disc.

The uncertainty in the parallax measurements of GSC\,08350-01924\,A and GSC\,08350-01924\,B allows the possibility that they have a binary separation between 25 and 135 au, but the separation could also be more than 135 au. The multiplicity of these stars may or may not affect the likelihood of the presence of a detectable debris disc. The on-sky separation of GSC\,08350-01924\,AB is less than the observation beam size.

\subsection{Non-significant ALMA excesses}

We now turn to the observations starting with a few systems that do not have a significant excess, but were close enough to warrant further investigation. The list of non-detections can be obtained from Table \ref{tab:ObsDet}, i.e. the sources that are not marked in boldface.

\subsubsection{TYC\,7443-1102-1}\label{sec:TYC7443}

\begin{figure}
        {\includegraphics[width=1\columnwidth]
        {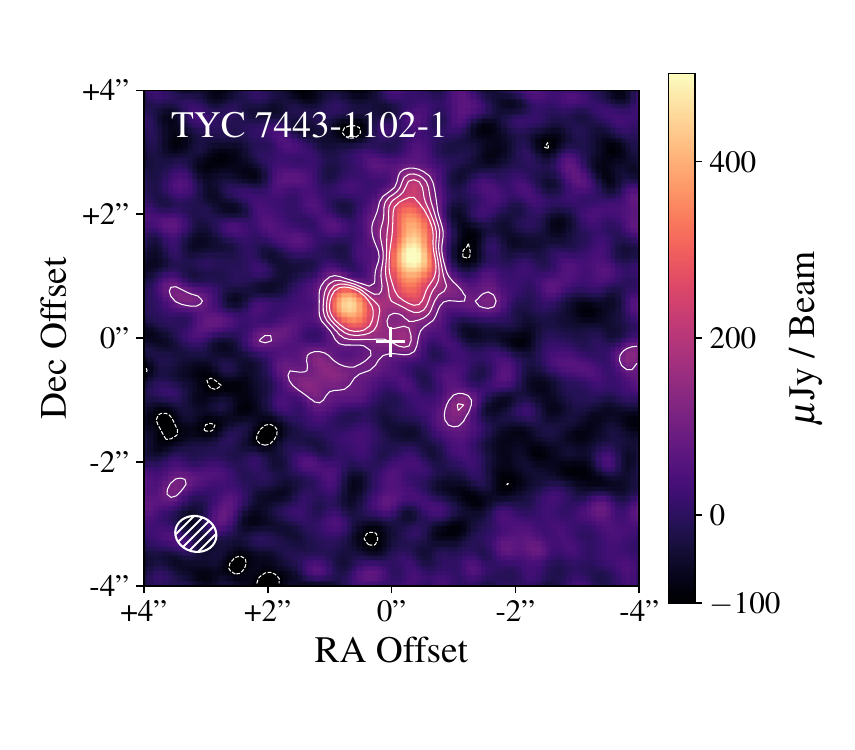}}
        \caption[Naturally weighted ALMA image of TYC\,7443-1102-1.]{\label{fig:TYC7443} Naturally weighted ALMA 880$\mu$m image of TYC\,7443-1102-1. The stellar location is marked with a +. The ellipse in the lower left corner shows the restoring beam. Contours are -3$\sigma$, -2$\sigma$, 2$\sigma$, 3$\sigma$, 4$\sigma$, 5$\sigma$. }
\end{figure}

This star has an unresolved Herschel PACS excess as reported in \citet{Tanner20}.
Two distinct sub-mm sources are clearly detected in the ALMA observation displayed in Figure \ref{fig:TYC7443}, neither of which are centred at the Gaia DR2 proper-motion adjusted location of the star. The two sources are 1.4" and 0.9" distant from the stellar location and have integrated flux densities of 2.20$\pm$0.05\,mJy and 0.44$\pm$0.05\,mJy respectively. The brighter of the two sources is resolved along one axis.

The ALMA absolute pointing accuracy for this observation is $\sim$30 mas and the error on the Gaia stellar location is sub-milliarcsecond, and so the separation of the sources from the expected stellar location is most likely accurate. The flux of these sources are not inconsistent with the flux expected from a debris disc with a radius equal to their separation from the star. However, if these sources constitute a debris disc such a disc would be more asymmetric than any other observed disc with no other known discs showing similar features.
Therefore, we conclude that these mm-wave sources are most likely not associated with the star and constitute background galaxies. 

For a putative debris disk to be detected with Herschel PACS but not with ALMA, the spectral slope of the dust emission would need to have $\beta \gtrsim 1$, where the dust emission is described a modified blackbody $F_{\nu} \propto B_{\nu}(\nu,T)(\frac{\lambda_0}{\lambda})^{\beta}$ beyond a turnover wavelength $\lambda_0$. This would be steeper than is seen for well-characterised cases \citep[e.g.][]{Gaspar12,MacGregor16}.
Larger surveys (that are less precise) find $\beta$ values in the range of $0.5$\,-\,$1$ \citep{Holland17,Sibthorpe18}. Thus a scenario where the PACS detection is of a circumstellar disk that is then not detected by ALMA is improbable. Therefore the Herschel excess most likely also originated from these contaminating sources and the conclusion is drawn that a circumstellar disk around TYC\,7443-1102-1 is not detected.

As the observation is significantly contaminated at the stellar location we remove the observation and star from the scientific sample going forward.

\subsubsection{HD\,155555\,C}

\begin{figure}
        {\includegraphics[width=1\columnwidth]
        {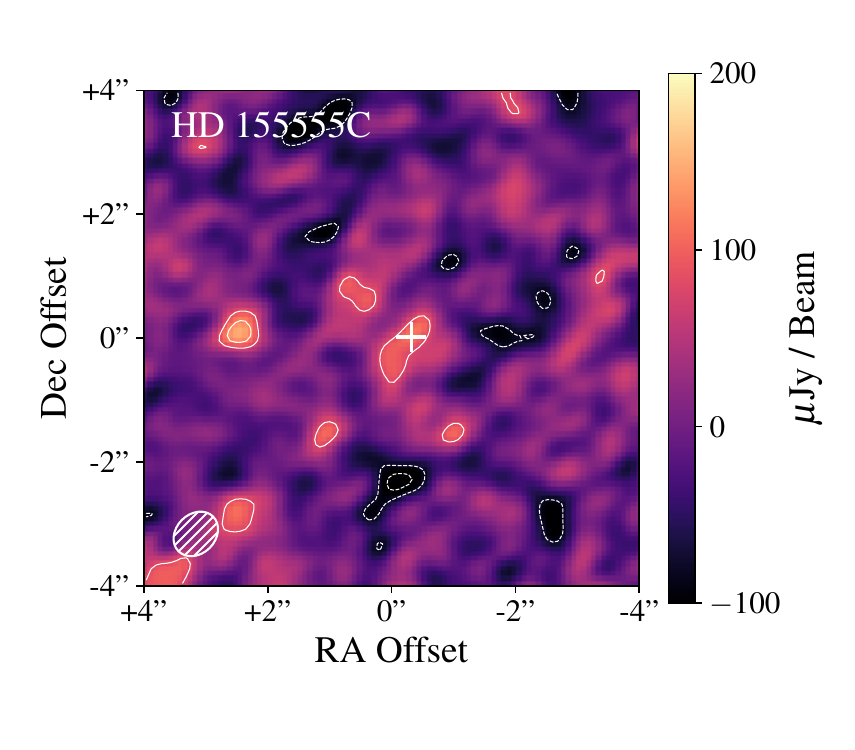}}
        \caption[Naturally weighted ALMA image of HD\,155555\,C.]{\label{fig:HD15C} Naturally weighted ALMA 880$\mu$m image of HD\,155555\,C. The stellar location is marked with a +. The ellipse in the lower left corner shows the restoring beam. Contours are -3$\sigma$, -2$\sigma$, 2$\sigma$, 3$\sigma$, 4$\sigma$, 5$\sigma$.}
\end{figure}

The 93$\pm$40\,$\mu$Jy\,beam$^{-1}$ flux at the stellar location of this observation, as displayed in Figure \ref{fig:HD15C}, is between 2$\sigma$ and 3$\sigma$, and so it warranted a further analysis. We apply the \textit{uvmodelfit} task again, now allowing the offset parameters to vary, and find a flux of 116\,$\pm$40$\mu$Jy\,beam$^{-1}$ at a separation of 0.21$\pm$0.07\,\arcsec, that could be consistent with the stellar location.

The stellar flux is only expected to be 9\,$\mu$Jy\,beam$^{-1}$ and so if the flux is real it would constitute an excess. As there are multiple 2$\sigma$ peaks within 2\,\arcsec\,of the stellar location, combined with the offset of the flux, we rule the flux measurement to likely be the result of noise. Given 33 observations there is approximately a 10\% chance that at least one observation will have a 3$\sigma$ peak at the stellar location. Given that HD\,155555\,C is the only source in our sample with a near-detection, we consider it likely that the excess flux in this observation is simply noise and the result of observing a moderately large number of systems. However this star is still worth re-observing in order to discover or rule out the presence of an infrared excess with more significant certainty.

\subsubsection{AT\,Mic\,B}%

\begin{figure}
        {\includegraphics[width=1\columnwidth]
        {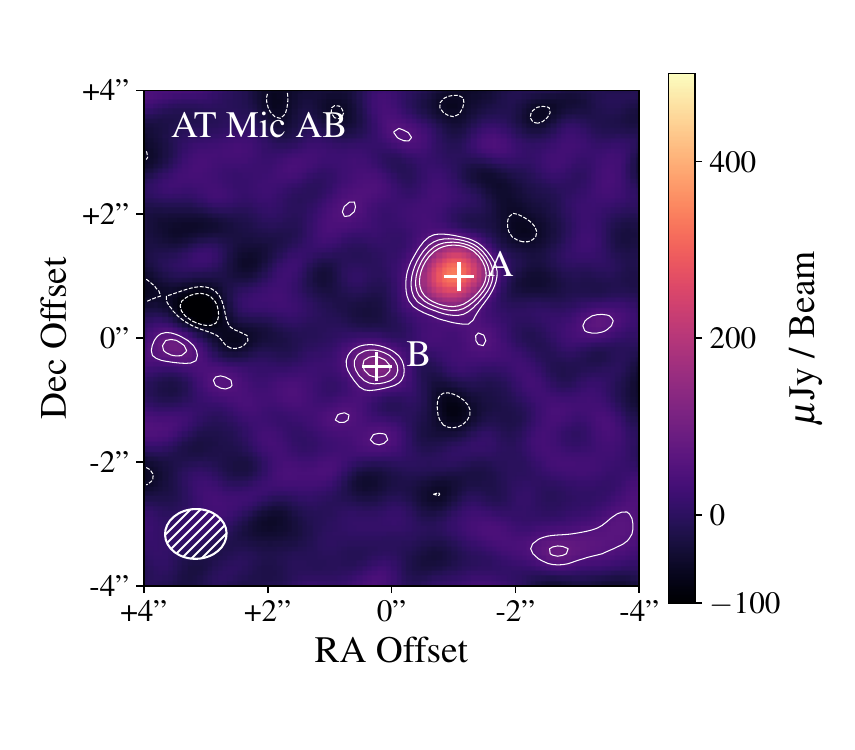}}
        \caption[Naturally weighted ALMA image of AT\,Mic\,AB.]{\label{fig:ATMic} Naturally weighted ALMA 880$\mu$m image of AT\,Mic\,AB. The stellar locations are marked with a + and an A/B. The ellipse in the lower left corner shows the restoring beam. Contours are -3$\sigma$, -2$\sigma$, 2$\sigma$, 3$\sigma$, 4$\sigma$, 5$\sigma$. }
\end{figure}

A flux of 120$\pm$27\,$\mu$Jy\,beam$^{-1}$ is measured at the stellar location of this observation, as displayed in Figure \ref{fig:ATMic}, reaching a significance of 4$\sigma$. We apply the \textit{uvmodelfit} task again, now allowing the offset parameters to vary, and find a flux of 125\,$\pm$27$\mu$Jy\,beam$^{-1}$ at a separation of 0.09$\pm$0.06\,\arcsec, consistent with the expected Gaia DR3 stellar location.

However, the expected stellar flux is 60\,$\mu$Jy\,beam$^{-1}$. The star is therefore confidently detected, but after subtracting the expected stellar flux the remaining mm-wave excess of $65 \pm 27$\,$\mu$Jy\,beam$^{-1}$ does not reach 3$\sigma$ for this observation. And so we conclude that an excess is not significantly measured for this star.

\subsection{Significant ALMA excesses}

\begin{figure}
        \includegraphics[width=1\columnwidth]{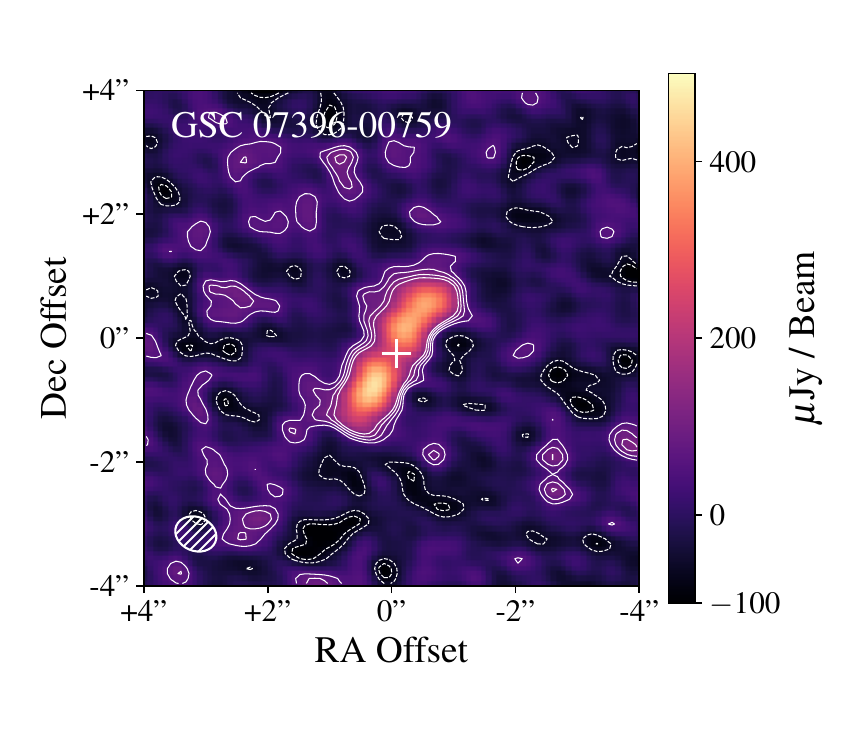}
        \vfill
        \includegraphics[width=1\columnwidth]{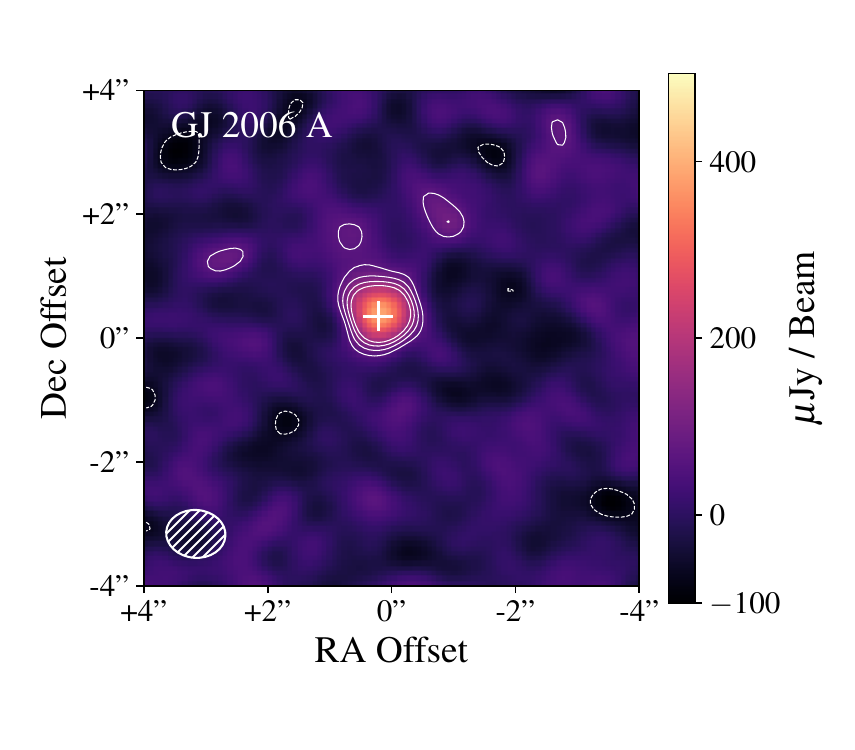}
        \vfill
        \includegraphics[width=1\columnwidth]{ATMicImage.pdf}
        
        \caption[Naturally weighted ALMA images of GSC\,07396-00759, GJ\,2006\,A and AT\,Mic\,AB.]{\label{fig:Dets} Naturally weighted ALMA 880$\mu$m images of GSC\,07396-00759, GJ\,2006\,A and AT\,Mic\,AB. The stellar locations are marked with a +. The ellipses in the lower left corners show the restoring beams. Contours are -3$\sigma$, -2$\sigma$, 2$\sigma$, 3$\sigma$, 4$\sigma$, 5$\sigma$.}
\end{figure}

\begin{figure}
        \includegraphics[width=1\columnwidth]{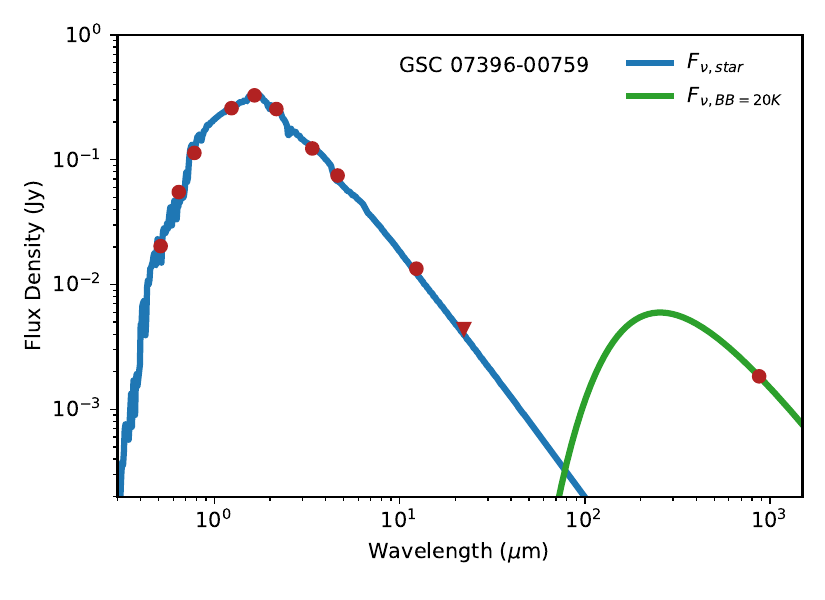}
        \vfill
        \includegraphics[width=1\columnwidth]{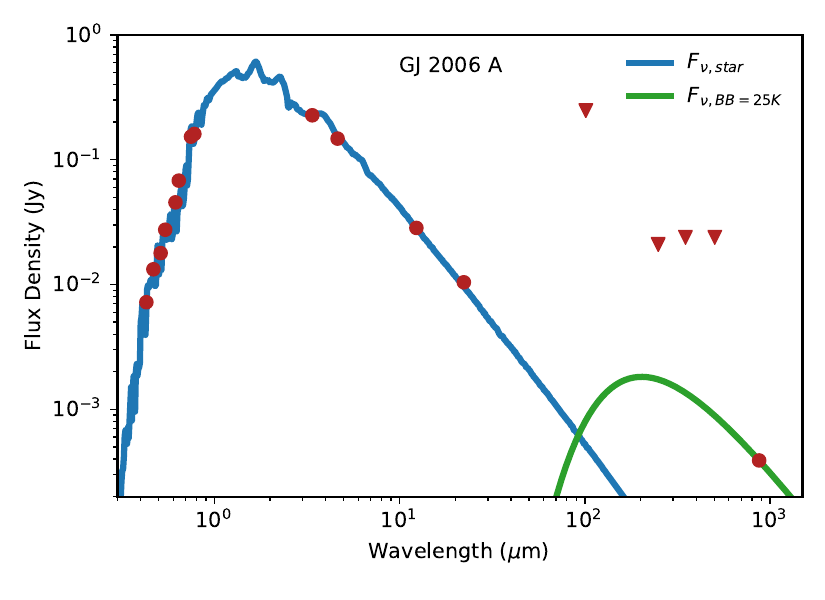}
        \vfill
        \includegraphics[width=1\columnwidth]{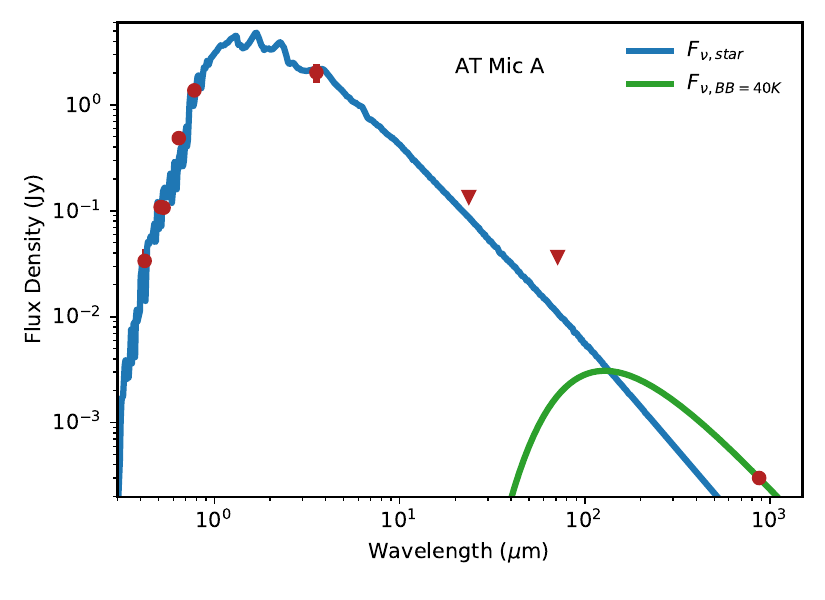}
        
        \caption[Example SEDs for GSC\,07396-00759, GJ\,2006\,A and AT\,Mic\,AB.]{\label{fig:seds} Example SEDs for GSC\,07396-00759, GJ\,2006\,A and AT\,Mic\,A. Dots are measured fluxes and triangles are 3$\sigma$ upper limits. The stellar photosphere models are in blue and example blackbody distributions through the ALMA flux are in green. With only one flux point measuring the thermal emission of the discs, a large range of temperatures and fractional luminosities could describe the discs.}
\end{figure}

\begin{figure}
        \includegraphics[width=1\columnwidth]{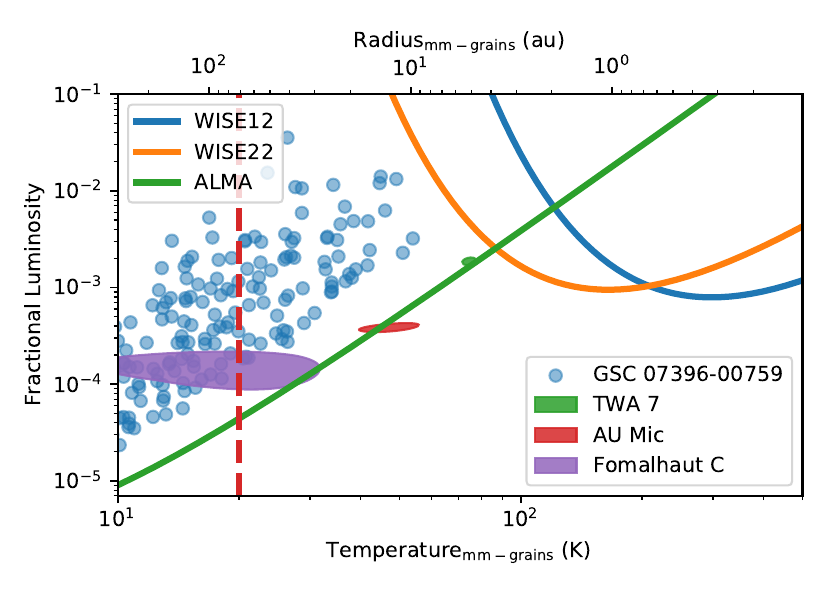}
        \vfill
        \includegraphics[width=1\columnwidth]{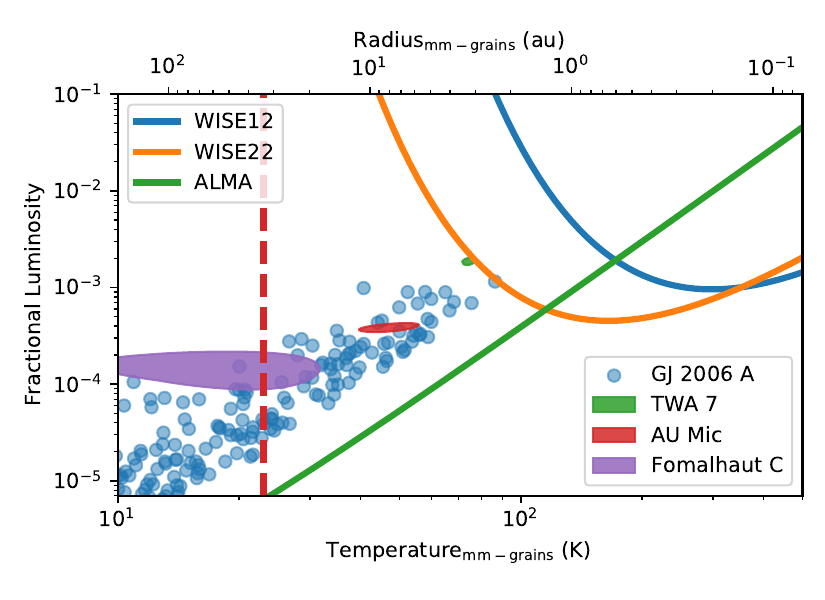}
        \vfill
        \includegraphics[width=1\columnwidth]{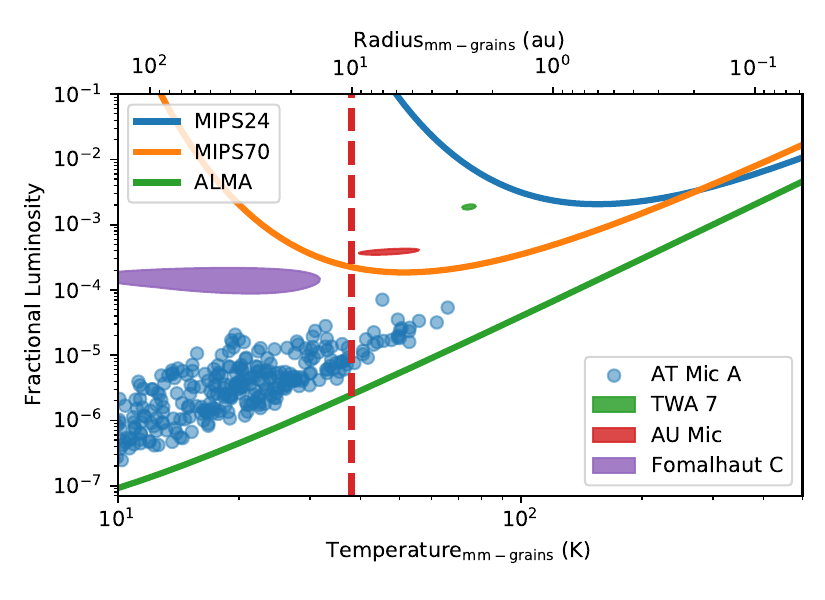}
        
        \caption[Fractional luminosity - temperature plots for GSC\,07396-00759, GJ\,2006\,A and AT\,Mic\,AB.]{\label{fig:fts} Plots of fractional luminosity against representative temperature/blackbody radius, i.e. the temperature and stellocentric radius of mm grains. Blackbody radius depends on host stellar temperature and is thus only accurate for the host of interest. A selection of allowed models for the discs of GSC\,07396-00759, GJ\,2006\,A and AT\,Mic\,A are plotted as blue circles. The distributions up to 3$\sigma$ following the same modified blackbody SED fitting procedure are shown for a selection of low mass host debris discs as coloured ellipses. The detection limits for several instruments are plotted as blue, orange and green curves respectively. The vertical red dashed lines show the resolved radius or radius upper limits of the discs.}
\end{figure}

\subsubsection{GSC\,07396-00759}

This observation clearly resolves a bright, edge-on debris disc, as displayed in Figure \ref{fig:Dets} with position angle, inclination and approximate radius consistent with the previous scattered light observations of this disc \citet{Sissa18,Adam21}. An in-depth analysis of the ALMA data for this disc is presented in \citet[][]{Coltsmann22}. 

The disc has an integrated mm flux of 1.84$\pm$0.22\,mJy and a radius of 70.2$\pm$4.4\,au, an example SED is displayed in Figure \ref{fig:seds} and a fractional luminosity-temperature plot with a distribution of dust models is displayed in Figure \ref{fig:fts}. 
The fractional luminosity-temperature plot shows different fitted models of the disc's fractional luminosity and the temperature of its mm-dust grains, which is related to the radial distance of those grains from the star.
These models must be compatible with the SED of the disc but do not take into account resolution effects or radial information derived from the image of the disc.
In comparison to these models are displayed models of other well characterised M-dwarf discs and the detection limits of several relevant mid-to-far-infrared instruments. The plot also shows the radius of the disc as observed by ALMA.
With a lack of far-IR photometry it is difficult to constrain an SED and model temperature, but with a resolved radius of 70.2\,au the mm dust grains would have a temperature of 20 K and so we can limit the likely models to those close to 20\,K, i.e. close to the dashed red line in Figure \ref{fig:fts}. Limited to these models, the fractional luminosity likely ranges from $\sim 1\times10^{-4}$-$5\times10^{-3}$. More details on the SED fitting procedure can be found in \citet{Yelverton19}.

\subsubsection{GJ\,2006\,A}

A flux of 390$\pm$33\,$\mu$Jy\,beam$^{-1}$ is measured at the stellar location of this observation, as displayed in Figure \ref{fig:Dets}, reaching a significance of 11$\sigma$. We apply the \textit{uvmodelfit} task again, now allowing the offset parameters to vary, and find a flux of 391$\pm$27$\mu$Jy\,beam$^{-1}$ at a separation of 0.03$\pm$0.02\,\arcsec, consistent with the expected Gaia DR3 stellar location. Subtracting the expected stellar flux of 6\,$\mu$Jy\,beam$^{-1}$ from the measured flux leaves a mm excess of 385$\pm$33$\,\mu$Jy\,beam$^{-1}$, remaining at 11$\sigma$.

Having ruled out stellar flaring this mm excess likely constitutes an unresolved debris disc. The beam size of the observation sets an upper limit on the radius of the disc, a beam semi-major axis of 0.96\,\arcsec\,sets a radius upper limit of 34 au. 
An example SED is presented in Figure \ref{fig:seds} and a fractional luminosity-temperature plot with a distribution of dust models is displayed in Figure \ref{fig:fts}.
The fractional luminosity-temperature plot shows the upper limit on the radius of the disc as observed by ALMA.
With a lack of far-IR photometry it is difficult to constrain an SED and model temperature, but with an upper limit of 34\,au on the disc radius we can place a lower limit on the mm grain temperature of 25 K, i.e. to the right of the dashed red line in Figure \ref{fig:fts}. Limited to these models the fractional luminosity likely ranges from $\sim$2$\times10^{-5}$-$1\times10^{-3}$. 

\subsubsection{AT\,Mic\,A}

A flux of 319$\pm$27\,$\mu$Jy\,beam$^{-1}$ is measured at the stellar location of this observation, as displayed in Figure \ref{fig:Dets}, reaching a significance of 11$\sigma$. We apply the \textit{uvmodelfit} task again, now allowing the offset parameters to vary, and find a flux of 335$\pm$27\,$\mu$Jy\,beam$^{-1}$ at a separation of 0.13$\pm$0.03\,\arcsec. 
Subtracting the expected stellar flux of 70\,$\mu$Jy\,beam$^{-1}$ from the measured flux leaves a mm excess of 265$\pm$27$\,\mu$Jy\,beam$^{-1}$, reaching 8$\sigma$.

We consider the apparent $\sim$0.13$\pm$0.03\,\arcsec\,separation, approximately one eighth of the beam size, between the expected stellar location of AT\,Mic\,A and the mm source. The uncertainty of the \textit{uvmodelfit} is not consistent with the stellar location; however, while Gaia positional astrometric uncertainties are reported as sub-milliarcsecond, the ALMA astrometric precision for this observation (calculated per \S10.5.2 of the ALMA Cycle 6 Technical Handbook\footnote{https://almascience.nrao.edu/documents-and-tools/cycle6/alma-technical-handbook}) is 0.065\,\arcsec. Considering also the 0.09$\pm$0.06\,\arcsec\,offset for AT\,Mic\,B's flux, which is in a similar direction, it is likely that the offset for both stars is the result of either uncertain ALMA pointing or possibly the effect of orbital motion. Having also ruled out stellar flaring, we conclude that this excess flux is evidence of an unresolved debris disc around AT\,Mic\,A.

The beam size of the observation sets an upper limit on the radius of the disc: a beam semi-major axis of 1\,\arcsec\,sets a radius upper limit of 10 au. The semi-major axis of \citet{Malkov12} of 31\,au would make this disc the first binary system to have a detected debris disc where the binary separation is between 25 and 135\,au, however it is uncertain if \citet{Yelverton19}'s conclusions extend to M dwarfs and if not, this may not be unusual. 

An example SED is presented in Figure\,\ref{fig:seds} and a fractional luminosity-temperature plot with a distribution of dust models is displayed in Figure\,\ref{fig:fts}.
The fractional luminosity-temperature plot shows the upper limit on the radius of the disc as observed by ALMA.
With a lack of far-IR photometry it is difficult to constrain an SED and model temperature, but with an upper limit of 10\,au on the disc radius we can place a lower limit on the mm grain temperature of 40\,K, i.e. to the right of the dashed red line in Figure \ref{fig:fts}. Limited to these models the fractional luminosity likely ranges from $\sim$5$\times10^{-6}$-$5\times10^{-5}$.  

\section{Discussion}\label{sec:Disc}

\subsection{Survey sensitivity and detection fraction}

To review our BPMG M-dwarf sample, excluding TYC\,7443-1102-1 and including AU Mic, we have: 33 observations containing 34 well resolved and well separated literature M dwarfs; an additional three Gaia DR3 M dwarfs with parallaxes (2MASS\,J05241914-1601153B, LP\,476-207B, GSC\,08350-01924B), although one of these three stars is close enough to the primary that a disc would likely be circumbinary (2MASS\,J05241914-1601153B); two of the total sample stars are also spectroscopic binaries without resolved companions (Barta 161 12, TYC\,6872-1011-1); and there are an additional 2 Gaia DR3 M dwarf candidates without parallaxes (potential companions to 2MASS\,J19102820-2319486, UCAC3\,124-580676). We treat binaries where dust is likely circumbinary as one system for the sake of the sample, and we do not include stars without Gaia DR3 parallaxes as we cannot verify that they are local M-dwarfs and not more distant brighter stars. With these constraints our scientific sample is 36 M-dwarf hosts. 

Of these systems we have four significant detections, GSC\,07396-00759, GJ\,2006\,A, AT\,Mic\,A and AU\,Mic. This makes our detection rate 4/36 or 11.1\%. We derive an uncertainty on this using the uncertainty in small number binomial statistics method set out in the appendix of \citet{Burgasser03}, for a result with uncertainties of 11.1$^{+7.4}_{-3.3}$\%.

\begin{figure}
        {\includegraphics[width=1\columnwidth]
        {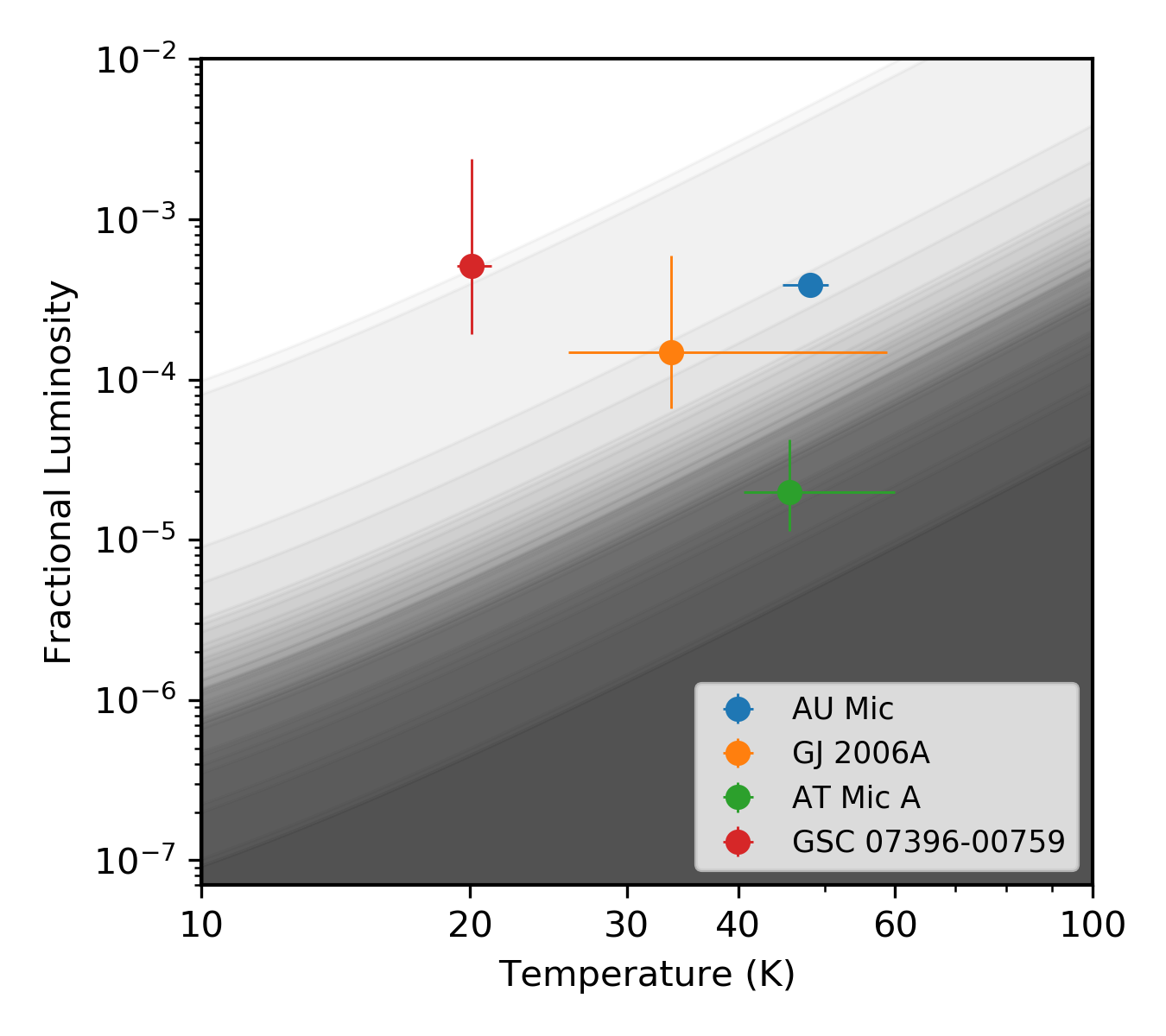}}
        \caption[Plot of detection limits for all observations.]{\label{fig:Completeness} Plot of detection limits for all observations. Local shading shows completeness for that observation where the white in the top-left corner is 100\% and the darkest gray in the bottom right corner is 0\%. The four detections of our sample are plotted to demonstrate their completeness and the survey's general sensitivity. }
\end{figure}

We can also calculate a completeness adjusted detection rate, adjusting for the survey's differing sensitivity for different observations. This is calculated by measuring the completeness for each of our detections, i.e. if that disc flux were present for each observation, what fraction of the observations would have significantly detected it? This is exemplified in Figure\,\ref{fig:Completeness}, in which the shading indicates the local completeness. In the dark bottom of the plot no observation would have been able to detect a disc, and in the white top all observations would have been able to detect a disc. We have plotted our four detections with 1$\sigma$ error bars from the fractional luminosity-temperature distributions seen in Figure \ref{fig:fts}, after constraining them with our disc radius information. For GSC\,07396-00759 only the models with a disc radius within 4.3\,au of 70.2\,au are considered, in accordance with the radius fitting of \citet[][]{Coltsmann22}; only the models with a disc radius smaller than 34\,au and 10\,au are considered for GJ\,2006\,A and AT\,Mic\,A respectively.
The completeness fraction for our four sources are: GSC\,07396-00759: 36/36, i.e. all our observations could have detected a GSC\,07396-00759-like disc if one were present; GJ\,2006\,A: 33/36; and AU\,Mic: 33/36; AT\,Mic\,A: 7/36, i.e. only seven of our observations were sensitive enough to have detected an AT\,Mic\,A-like disc. Dividing through by these completion fractions and summing results in our completeness adjusted detection fraction: 8.3/36 or 23.1\%. With the same method of uncertainties applied we get: 23.1$^{+8.3}_{-5.3}$\%.

Given that much of the weight of this completeness adjusted result derives from AT\,Mic\,A alone, an effect that is exacerbated in the small number regime, and as the uncertainties in the disc parameters are not taken into account, the uncertainties on the completeness adjusted detection rate are likely underestimated. To investigate these effects we generated one million sets of four synthetic debris disc detections; we chose sets of four synthetic detections as there were four real detections within our sample. Within each set each disc had a radius selected randomly from between 10 and 100\,au with linearly spaced probability and a fractional luminosity selected randomly from between $10^{-3}$ and $10^{-7}$ with logarithimically spaced probability. The host star luminosity was then selected randomly from the luminosities of the stars in our sample without replacement. The completeness adjusted detection rate was calculated for each set and over the one million sets a distribution of synthetic completeness adjusted fractions was formed. The median of this distribution with its distance to the 16th and 84th percentiles was 29.9$^{+12.3}_{-8.9}$\%.
While the synthetic rate is not significantly larger than the observed fraction, its greater uncertainty does imply that the uncertainties on the observed completeness adjusted detection rate are indeed likely underestimated. 
This process has made large assumptions about the underlying M-dwarf disc population, however there are not yet enough well-observed M-dwarf debris discs to build a more informed model population.

The completeness adjusted detection rate implies that there could be another four AT\,Mic\,A-like discs hiding amidst the rest of the sample but that the observations were not sensitive enough to detect them.

\subsection{Detection rate in context}

To begin with, we compare our 11.1$^{+7.4}_{-3.3}$\% detection rate and 23.1$^{+8.3}_{-5.3}$\% completeness adjusted rate to the DEBRIS M sample. The DEBRIS survey detected just 2/89 (2.2$^{+3.4}_{-2.0}$\%) M-dwarf discs; immediately our detection rate is significantly higher. However, we cannot conclude that this is due to ALMA's capability to detect M-dwarf discs over Herschel's, as \citet{Pawellek21} measure a 9/12 (75\%) detection rate for F star discs in the BPMG, compared to the 22/92 (23.9$^{+5.3}_{-4.7}$\%) rate for F stars of the DEBRIS survey presented in \citet{Sibthorpe18}. If whatever was the root cause of \citet{Pawellek21}'s high detection rate for BPMG F stars holds for BPMG M stars, be it a matter of youth, formation environment or some other factor, it could raise the base detection rate. In a simple calculation, if the BPMG has an approximately three times higher detection base rate, the DEBRIS M-dwarf rate adjusted to the BPMG M-dwarf sample would only be 6\%, still nearly half our non-adjusted rate, although within uncertainty due to the small number statistics. Comparing also to the 1/900 detection rate of \citet{Rhee07}'s IRAS search for M-dwarf discs and \citet{Gautier07}'s 0/62 Spitzer detection rate, we do conclude that ALMA has enabled us to probe M-dwarf discs in a way that previous telescopes were not able to due to their wavelength and sensitivity limitations.

Comparing our M-dwarf BPMG sample to \citet{Pawellek21}'s F-type BPMG sample, our detection rate is seven times lower than the F-type rate. However, the F-type sample are all within 25\,pc, unlike our M-type sample that ranges up to 100\,pc. To account for this we should compare our completeness-corrected rate, but this is still three times lower. F-types have been previously measured to possess greater detection rates than G and K types, but only by a factor of $\sim$1.7 as measured by \citet{Sibthorpe18} in the DEBRIS FGK sample.
It is possible that the higher rate arises because brighter host stars illuminate the discs more, allowing them to be more easily detected. \citet{Pawellek21}'s sample ranges from F0V to F9V (5.71 $L_{\odot}$ to 1.69 $L_{\odot}$) while our M-type sample ranges from M0V - M8.5V (0.275 $L_{\odot}$ to 0.004 $L_{\odot}$). M-dwarf samples span a large luminosity range and their luminosities can be several orders of magnitude lower than FGK type star luminosities. It is possible that the F-type BPMG sample and the M-type BPMG sample host similar discs but the host luminosities affect observability too significantly. That is, while ALMA provides an increase in sensitivity over previous far-IR observations for discs around M-type stars, it may still be that far-IR observations of earlier type stars yield a higher detection rate than ALMA observations of later type stars.
It is also possible that whatever mechanism boosts the detectability of BPMG F-type discs does not apply to late type stars; this scenario would mean we can more directly compare our results to age-spread field star surveys like DEBRIS.

Compared to the Herschel DEBRIS G and K samples' detection rates of 14.3$^{+4.7}_{-3.8}$\% and 13.0$^{+4.5}_{-3.6}$ respectively and completeness adjusted rates of 24.6$^{+5.3}_{-4.9}$\% and 22.5$^{+5.6}_{-4.2}$ respectively, our 11.1$^{+7.4}_{-3.3}$\% detection rate and 23.1$^{+8.3}_{-5.3}$\% completeness adjusted rate are consistent, if not following the slight trend of decreasing detection rate with type. This similarity suggests that the difference between our sample and the BPMG F-types is more related to an unusual property of those F-type stars than a large difference in ALMA versus far-IR sensitivity as a function of spectral type.

We now compare to the \citet{Luppe20} predictions for an ALMA survey of DEBRIS-like M-dwarf discs. Our sample has been observed for approximately 15 minutes per star with ALMA Band 7, and the observations were designed to reduce the likelihood that discs would be resolved. It is unlikely that any discs would be larger than the maximum recoverable scales of our observations, but as evidenced by GSC\,07396-00759 discs could still have been resolved, reducing the flux per beam. Without correcting for resolution \citet{Luppe20} predict 15 minutes of observation at Band 7 of the Herschel DEBRIS sample of M-dwarfs scaled as DEBRIS-like discs to attain a detection rate of 4.3$\pm$0.9\% to 15.8$\pm$0.5\%, entirely consistent with our observations. If the DEBRIS sample and the BPMG stellar samples are broadly similar, this would imply that M-dwarf discs are overall similar to earlier type stars' discs in terms of radius, total surface area, temperature and fractional luminosity, when scaled by stellar mass and luminosity. 

The DEBRIS sample is selected from the closest stars, but over a range of ages. \citet{Pawellek21} has shown based on their high detection rate for F type discs that the BPMG sample could be significantly different to the DEBRIS sample. Ultimately, to investigate whether M-dwarf discs differ from earlier type discs one would need to use the scaling relationships of \citet{Luppe20} and apply their process to the known FGK-type BPMG discs to produce a theoretical FGK-like M-dwarf sample to compare our sample to. However, the small number statistics would likely inhibit differentiation of \citet{Luppe20}'s different scaling relationships. 

Ultimately we conclude that our ALMA Band 7 detection rate is evidence that M-dwarf discs are not significantly less common than earlier type discs, but that the telescopes employed in previous surveys could not efficiently observe the low temperature and fluxes of M-dwarf discs due to their low host luminosities.

\subsection{Radii in context}

\begin{figure}
        {\includegraphics[width=1\columnwidth]
        {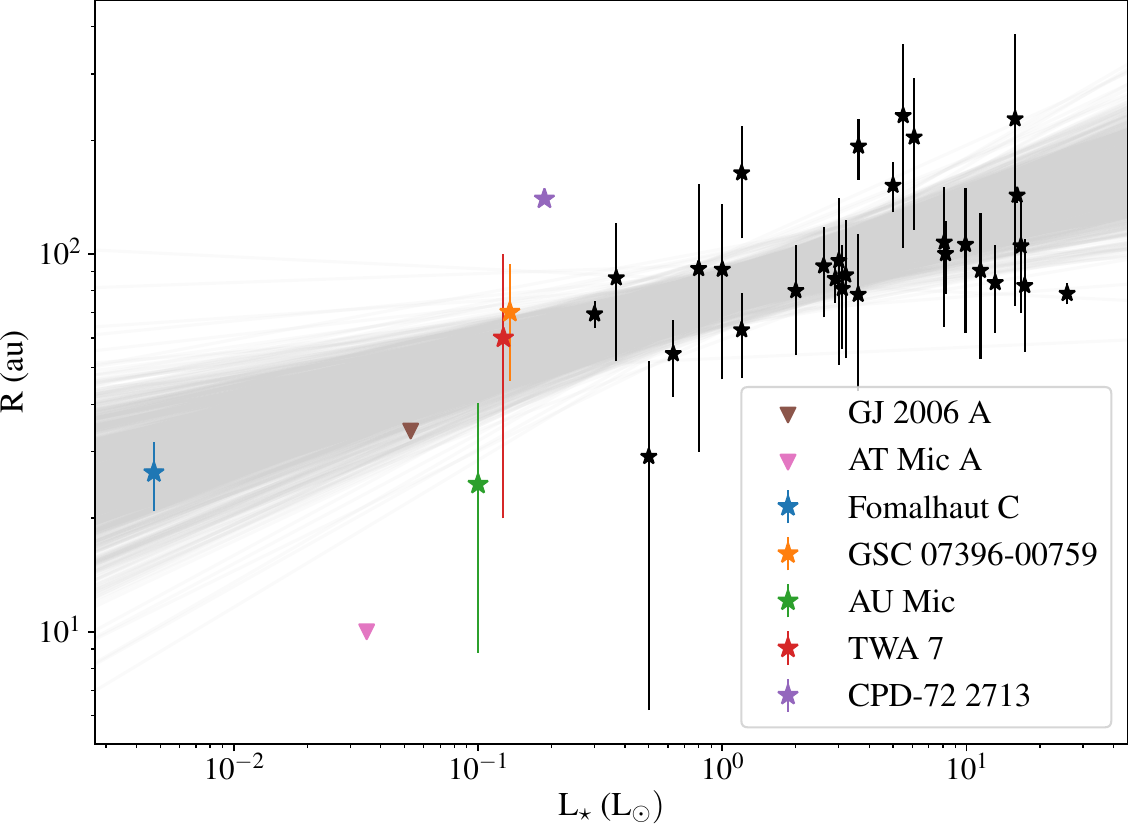}}
        \caption[mm-wave resolved debris disc radii plotted against host stellar luminosity.]{\label{fig:LRPlotBPMG} mm-wave resolved debris disc radii plotted against host stellar luminosity. Error bars represent disc FWHM or upper limits. The five latest type stellar hosts are highlighted in colour, CPD-72\,2713 is plotted without a width  as a fixed width of 0.2R was assumed to facilitate fitting a radius \citep{Moor20}. Also added are upper limits for the discs of GJ\,2006\,A and AT\,Mic\,A. Transparent grey lines show a sample of 1000 power laws from the parameter distributions of \citet{Matra18}.}
\end{figure}

In Figure\,\ref{fig:LRPlotBPMG} we plot the mm-wave radii of all mm resolved debris discs against the host luminosity, as first presented in \citet{Matra18}; added to the original sample are the stars presented in \citet{Sepulveda19}, Fomalhaut\,C \citep{Coltsmann21} and CPD-72\,2713 \citep{Moor20}. We plot the resolved radius of the GSC\,07396-00759 disc and upper limits for GJ\,2006\,A and AT\,Mic\,A. We can see that GSC\,07396-00759's radius is consistent with the trend of the earlier type sample, if the disc of GJ\,2006\,A is close to the upper limit it would also be consistent. Although there is a large scatter, the upper limit on the radius of AT\,Mic\,A's disc is very small. However, we note that this is specifically a plot of resolved radii and that many discs of radii less than ten au have been inferred from SEDs, and they could not be resolved due to instrumental constraints, as this disc is not resolved due to instrument constraints. The AT\,Mic\,A disc would still be small by mm-wave detection standards, however the sample of discs at this low luminosity is small and it remains unknown whether this radius limit would be unusual for its host luminosity and mass. As the AT\,Mic binary are only separated by 30\,au, their orbits would prevent circumstellar discs larger than approximately 10\,au from surviving.

\section{Conclusion}

The Beta Pictoris Moving Group provides an excellent candidate sample of M-dwarfs to observe with ALMA to uncover new M-dwarf debris discs and resolve the question as to whether M-dwarf discs are rare or just difficult to detect. In this paper we have presented new ALMA Band 7 observations of 33 M dwarf systems comprising at least 37 M-dwarf stars. 

We identify one resolved disc, GSC\,07296-00759 with an integrated flux of 1.84\,mJy, and identify two unresolved mm-wave excess detections around GJ\,2006\,A with a flux of 385 $\mu$Jy\,beam$^{-1}$ and AT\,Mic\,A with a flux of 265 $\mu$Jy\,beam$^{-1}$.
We confirm that none of these stars show evidence of stellar flaring and none of the discs show evidence of $^{12}$CO J=3-2 emission. We explore the fractional luminosity-temperature parameter space for these discs and present fractional luminosity ranges.

We note two of our observations come close to our 3$\sigma$ criterion for detection. The flux at the stellar location of HD\,155555\,C could be noise or a dim excess, the star may be worth considering for future re-observation. AT\,Mic\,B has a 4$\sigma$ flux at the stellar location, but only a 2$\sigma$ excess above the expected stellar flux and so cannot be confirmed as a significant excess detection. This small excess, in addition to its proximity at 9.8 pc and its association with AT\,Mic\,A and AU Mic, makes this star worth re-observing in the future. If future observations of AT\,Mic\,A are made, AT\,Mic\,B will naturally be observed due to the small binary separation, and so it may be likely that this star's disc hosting candidacy will be determined in the future. 

We calculate a detection rate of 4/36, 11.1$^{+7.4}_{-3.3}$\%, for our M-dwarf sample including AU Mic. We also present a completeness fractional luminosity-temperature plot for our observations and calculate a completeness adjusted detection rate of 23.1$^{+8.3}_{-5.5}$\%, but we note that these errors are very likely to be underestimated.

We place our detection rate in context and conclude that it is consistent with the Herschel DEBRIS GK detection rate and the ALMA survey predictions of \citet{Luppe20}. We therefore conclude that M-dwarf debris discs are not significantly less common than earlier type discs but instead require longer wavelength and more sensitive observations to account for the low host luminosity.

We examine the disc radius upper limits of our new detections and conclude that GJ\,2006\,A is likely consistent with the wider luminosity-radius sample and trend. While the upper limit on the disc of AT\,Mic\,A is particularly small, it resides in too sparse a parameter space to be fully contextualised.

We examine the consequences of new Gaia DR3 astrometric information for the multiplicity of our sample. Due to their binarity we estimate that three of our systems are very unlikely to possess detectable discs due to their separation and that six to seven of our systems have a reduced likelihood of possessing detectable discs assuming that the results of \citet{Yelverton19} extend to M-dwarfs. Another 13 of our stars have binary companions that should not affect disc detection likelihood.

We stack 21 of our non-detection observations with the stars within 0.5\,\arcsec\,of the observation phase centre and calculate a 3$\sigma$ upper limit on the mean mm-wave excess of 24\,$\mu$Jy\,beam$^{-1}$ for those stars. 

Finally, we identify 11 background sources, likely sub-mm galaxies, of which one is resolved. The occurrence of background sources is consistent with the predictions of galaxy number count models \citep{Popping20}.
We identify the observation of TYC\,7443-1102-1 as severely contaminated by two of these background galaxies.

\section*{Acknowledgements}
We thank the referee for a careful report and valuable comments.
PFCC is supported by the University of Warwick.
GMK and SM are supported by the Royal Society as a Royal Society University Research Fellows. 
LM acknowledges funding from the Irish Research Council under grant IRCLA/2022/3788.
SM is supported by a Junior Research Fellowship from Jesus College, University of Cambridge.
For the purpose of open access, the author has applied a Creative Commons Attribution (CC-BY) licence to any Author Accepted Manuscript version arising from this submission.
This paper makes use of the following ALMA data: ADS/JAO.ALMA\#2017.1.01583.S. ALMA is a partnership of ESO (representing its member states), NSF (USA) and NINS (Japan), together with NRC (Canada), MOST and ASIAA (Taiwan), and KASI (Republic of Korea), in cooperation with the Republic of Chile. The Joint ALMA Observatory is operated by ESO, AUI/NRAO and NAOJ.
This work has made use of data from the European Space Agency (ESA) mission {\it Gaia} (\url{https://www.cosmos.esa.int/gaia}), processed by the {\it Gaia} Data Processing and Analysis Consortium (DPAC, \url{https://www.cosmos.esa.int/web/gaia/dpac/consortium}). Funding for the DPAC has been provided by national institutions, in particular the institutions participating in the {\it Gaia} Multilateral Agreement.
\section*{Data Availability}

The data underlying this article are available in http://almascience.nrao.edu/aq/, and can be accessed with ALMA project ID: 2017.1.01583.S.









\bsp	
\label{lastpage}
\end{document}